\documentclass[fleqn,usenatbib]{mnras}

\usepackage{newtxtext,newtxmath}

\usepackage[T1]{fontenc}
\usepackage{ae,aecompl}

\usepackage{graphicx}	
\usepackage{amsmath}	
\usepackage{amssymb}	
\usepackage{gensymb}
\usepackage{multirow}

\usepackage{xargs}                      
\usepackage[pdftex,dvipsnames]{xcolor}  
\usepackage[draft,colorinlistoftodos,prependcaption,textsize=small]{todonotes}   
\newcommandx{\unsure}[2][1=]{\todo[linecolor=red,backgroundcolor=red!25,bordercolor=red,#1]{#2}}
\newcommandx{\change}[2][1=]{\todo[linecolor=blue,backgroundcolor=blue!25,bordercolor=blue,#1]{#2}}
\newcommandx{\info}[2][1=]{\todo[linecolor=OliveGreen,backgroundcolor=OliveGreen!25,bordercolor=OliveGreen,#1]{#2}}
\newcommandx{\improvement}[2][1=]{\todo[linecolor=Plum,backgroundcolor=Plum!25,bordercolor=Plum,#1]{#2}}

\usepackage[final]{changes}
\definechangesauthor[name={JvR}, color=red]{jvr}

\hypersetup{draft}

\newcommand{\unsim}{\mathord{\sim}} 
\newcommand{\kms}{\,km\,s$^{-1}$} 
\newcommand{\Msun}{$\mathrm{M_{\sun}}$}
\newcommand{\Rsun}{$\mathrm{R_{\sun}}$}
\newcommand{\logg}{$\log{g}$}

\title[EL CVn-type binaries in PTF]{Discovery of 36 eclipsing EL CVn binaries found by the Palomar Transient Factory}

\author[J. van Roestel]
{J. van Roestel$^1$\thanks{j.vanroestel@astro.ru.nl},
T. Kupfer$^2$,
R. Ruiz-Carmona$^1$,
P.J. Groot$^1$,
T.A. Prince$^2$,
\newauthor
K. Burdge$^2$,
R. Laher$^3$,
D.L. Shupe$^4$,
E. Bellm$^5$
\\
$^1$Department of Astrophysics/IMAPP, Radboud University Nijmegen, P.O.Box 9010, 6500 GL, Nijmegen, NL \\
$^2$Cahill Center for Astronomy and  Astrophysics, California Institute of Technology, Pasadena, CA 91125, USA \\
$^3$Spitzer Science Center, California Institute of Technology, Pasadena, CA 91125, USA \\
$^4$Infrared Processing and Analysis Center, California Institute of Technology, Pasadena, CA 91125, USA \\
$^5$Department of Astronomy, University of Washington, Box 351580, Seattle, WA 98195, USA }

\date{Accepted XXX. Received YYY; in original form ZZZ}

\pubyear{2017}

\begin{document}
\label{firstpage}
\pagerange{\pageref{firstpage}--\pageref{lastpage}}
\maketitle

\begin{abstract}
We report the discovery and analysis of 36 new eclipsing EL CVn-type binaries, consisting of a core helium-composition pre-white dwarf and an early-type main-sequence companion, more than doubling the known population of these systems. We have used supervised machine learning methods to search 0.8\,million lightcurves from the Palomar Transient Factory, combined with SDSS, Pan-STARRS and 2MASS colours. The new systems range in orbital periods from 0.46 to 3.8\,d and in apparent brightness from $\unsim14$ to $16$\,mag in the PTF $R$ or $g^{\prime}$ filters. For twelve of the systems, we obtained radial velocity curves with the Intermediate Dispersion Spectrograph at the Isaac Newton Telescope. We modelled the lightcurves, radial velocity curves and spectral energy distributions to determine the system parameters. The radii (0.3--0.7\,\Rsun) and effective temperatures (8000--17000\,K) of the pre-He-WDs are consistent with stellar evolution models, but the masses (0.12--0.28\,\Msun) show more variance than models \added{have} predicted. This study shows that using machine learning techniques on large synoptic survey \replaced{data}{samples} is a powerful way to discover substantial samples of binary systems in short-lived evolutionary stages. 
\end{abstract}



\section{Introduction}\label{sec:intro}

EL CVn binaries are eclipsing binaries containing a low mass ($\unsim$0.15--0.33\,\Msun) pre-helium white dwarf (pre-He-WD) and an A/F-type main sequence star. The prototype system, EL CVn, is part of a sample of 17 EL CVn systems \citep{2014MNRAS.437.1681M} discovered by SWASP \citep{2006PASP..118.1407P} with magnitudes in the range of 9<$V$<13. All lightcurves show ``boxy", shallow eclipses ($\lesssim$ 0.1\,mag depth) with periods between $\unsim$0.5d and $\unsim$3\,d, and in \replaced{most}{some} cases ellipsoidal variation due to the deformation of the A/F-star. The low radial velocity amplitudes ($\unsim$15--30$\mathrm{\,km\,s^{-1}}$) of the primaries confirm the low mass nature of the pre-He-WDs.

A total of 10 EL CVn systems were found in the Kepler survey: 
KOI-74 \citep{2010ApJ...715...51V,2012MNRAS.422.2600B}; 
KOI-81 \citep{2010ApJ...715...51V,2015ApJ...806..155M}; 
KOI-1375 \citep{2011ApJ...728..139C}; 
KOI-1224 \citep{2012ApJ...748..115B}; 
KIC-9164561, KIC-10727668 \citep{2015ApJ...803...82R};
KIC-4169521, KOI-3818, KIC-2851474, and KIC-9285587 \citep{2015ApJ...815...26F}. All these systems were studied in great detail, and by modelling the Kepler lightcurves in combination with radial velocity curves, all system parameters have been determined. Four of these systems contain small pre-He-WDs ($<0.05$\,\Rsun) and as a consequence their lightcurves feature shallow eclipses only detectable from space. The fact that 10 EL CVn-like systems are found in the Kepler field suggests that there should be many more in our Galaxy, in line with an estimate of the local space density from stellar evolution and population synthesis models, $4$--$10\times10^{-6}\,\mathrm{pc^{-3}}$ \citep{2017MNRAS.467.1874C}.

Besides the samples found by Kepler and SWASP, there were serendipitous discoveries of binaries related to EL CVn systems. The star V209 in $\omega$ Cen is likely an EL CVn binary \citep{2007AJ....133.2457K}, but the primary does not seem to be a typical main sequence star: its mass is 0.95$\mathrm{M_\odot}$ but it has a temperature of 9370\,K. OGLE--BLG--RRLYR--02792 is an eclipsing binary \replaced{which contains a large pre-He-WD which seems to be pulsating like an RR-Lyrae star}{with a pre-He-WD and an RR-Lyrae star: a pulsating, evolved star possibly originating from an A-type main sequence star} \citep{2012Natur.484...75P}. A possible non-eclipsing variant of an EL CVn binary is the star Regulus ($\alpha$ Leo). \citet{Gies2008} and \citet{Rappaport2009} found that Regulus A is a single-lined spectroscopic binary with a period of $40$\,d, consisting of an A-type primary and a companion with a mass of >0.3\,\Msun, at the upper end of the pre-He-WDs mass range.

EL CVn binaries share many characteristics with a new type of binary: ``R CMa''-type binaries are Algol binaries with a bloated, low mass, donor \citep[e.g.][]{2011MNRAS.418.1764B,2016AJ....151...25L}. They are very similar to EL CVn systems, except that they are semi-detached, and therefore considered the progenitors of EL CVn systems. Two ``detached R CMa'' systems have been identified using Kepler photometry and are now considered to be newly born EL CVn binaries (KIC-10661783; \citealt {2013A&A...557A..79L}, and KIC-8262223; \citealt{2017ApJ...837..114G}). 

EL CVn systems are part of a larger family of binaries where one component of the binary is an extremely low mass white dwarf (ELMWD). The majority of ELMWD-containing binaries without a main-sequence companion have white dwarf or neutron star companions instead \citep[e.g.][]{1995MNRAS.275..828M,2005ASPC..328..357V}. In these systems, the ELMWD dominates the luminosity, making them identifiable with a single spectrum. The ELM survey \citep{2010ApJ...723.1072B} uses this approach and has been successful in finding many ELMWDs in binary white dwarf systems.

In this paper, we present system parameters for 36 new EL CVn systems, all eclipsing, discovered using the Palomar Transient Factory (PTF). In Section~\ref{sec:identification} we describe the identification of the systems using supervised machine learning classifiers. In Section~\ref{sec:spectra} we discuss the spectroscopic follow-up of 12 of the new systems. In Section~\ref{sec:analysis} we discuss the analysis of the lightcurves, spectra and spectral energy distributions, and we \replaced{present}{show} the results in Section~\ref{sec:results}. In Section~\ref{sec:discussion} we compare our results with theoretical prediction and compare our sample with already known EL CVn binaries. We end with a summary and conclusion in Section~\ref{sec:summary}.

\section{Target selection} \label{sec:identification}
\subsection{The Palomar Transient Factory}
The Palomar Transient Facility (PTF) used the 1.2\,m Oschin Telescope at Palomar Observatory with a mosaic camera consisting of 11 CCDs. The CCDs have 4Kx2K pixels and the camera has a pixel scale of 1.02$\arcsec$/pixel, giving it a total field of view of 7.26 square degrees. PTF uses an automated image processing pipeline which does bias and flatfield corrections, source finding and photometry. All data is automatically processed, see \citet{2009PASP..121.1334R,2009PASP..121.1395L} for further information.

\subsection{Data}
For all objects detected by PTF, lightcurves are automatically generated \citep[see][]{2014PASP..126..674L} and lightcurve statistics are calculated. These statistics include, among others, the mean, root-mean-square, percentiles, $\mathrm{\chi^2}$-statistic, see \citet{PTFstats} for a full list. These lightcurve statistics are based on the lightcurve features used in \citet{2011ApJ...733...10R,2012ApJ...744..192R}, that are useful to distinguish different type\added{s} of variable stars. Important to note is that we do not use features related to any periodicity in the lightcurve. This has a practical reason; it is very difficult to automatically obtain a reliable period for all the PTF lightcurves because they are sparsely sampled and span many years.

\begin{figure*}
\includegraphics{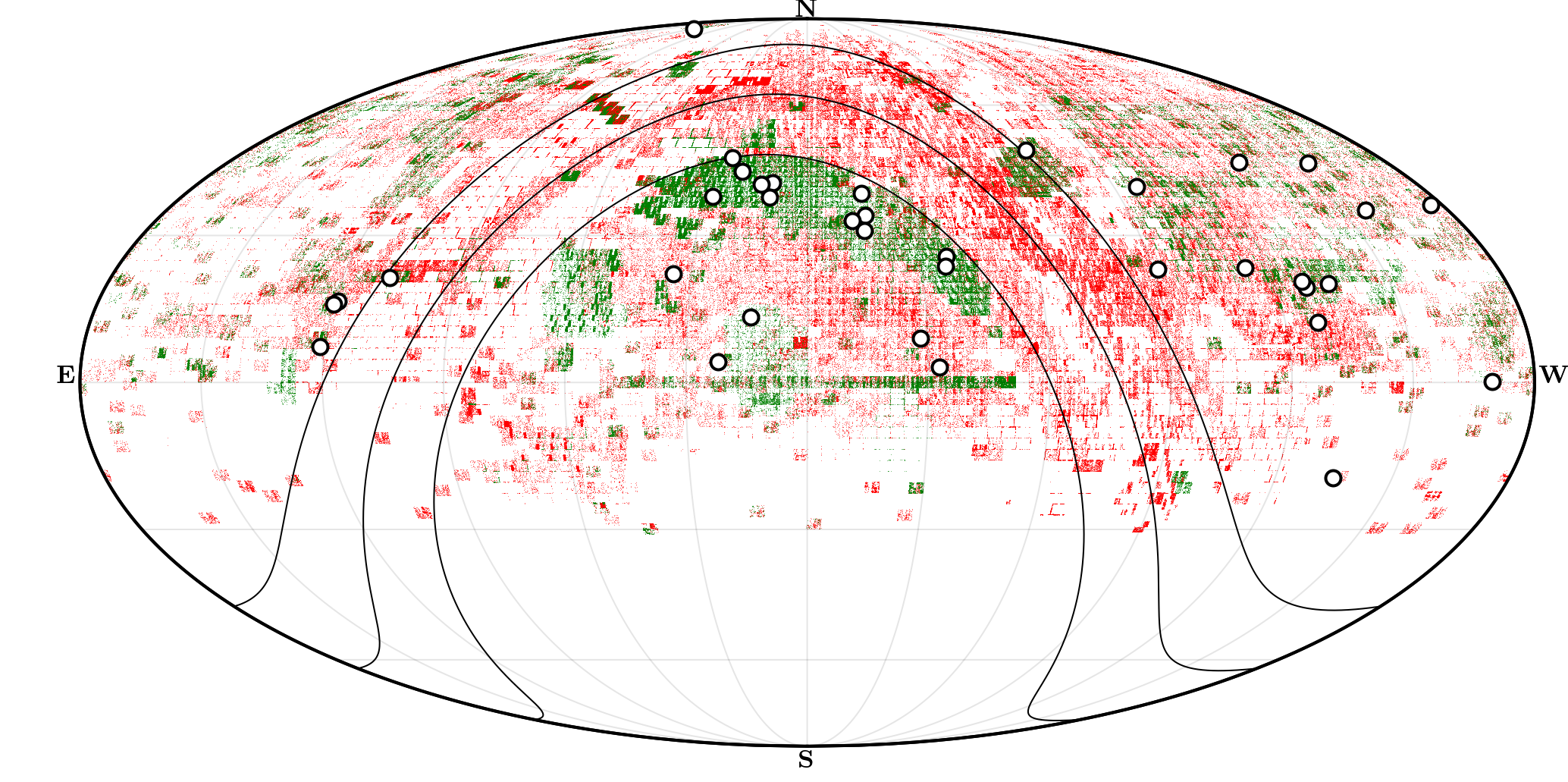}
\caption{All objects in the PTF sample after our initial cut ($>\!40$ epochs, $\chi^2_\mathrm{reduced} \! > \! 10$, $< \! 16$\,mag), red for PTF $R$, green for PTF $g^\prime$. The EL CVn binaries we discovered in the data are shown as white dots. The black lines show Galactic latitudes of $-15\degree$, $0\degree$, and $15\degree$.}
\label{fig:overview}
\end{figure*}

For this study, we used all available lightcurve data that was obtained between the start of PTF in December 2008 and March 2016. We treat the data for the $R$ and $g^\prime$ filter as two separate datasets in the subsequent analysis. These datasets are very substantial ($R$: $\unsim 250$ million, $g^\prime$: $\unsim 50$ million objects). We make an initial cut and select only objects which are variable by requiring that $\chi^2_\mathrm{reduced}>10$, that lightcurves have more than $40$ epochs, and that objects are brighter than 16\,mag in either $\mathrm{PTF}\,R$ or $\mathrm{PTF}\,g^\prime$. This still leaves more than $\unsim 10^5$ candidates (see Table~\ref{tab:PTFdata} and Fig.~\ref{fig:overview}).

We match the objects in these datasets to the latest SDSS catalogue \citep[$ugriz$ bands, DR13,][]{2016arXiv160802013S}, the NOMAD catalogue \citep[$JHK$ bands,][]{2004AAS...205.4815Z}, and the Pan-STARRS catalogue \citep[$grizy$ bands,][]{2016arXiv161205560C}. An overview of the total number\deleted{s} of objects and the colour availability are given in Table~\ref{tab:PTFdata}.

\begin{table}
\centering
\caption{The number of objects after our initial selection with PTF lightcurves (>40 epochs, $\chi^2_\mathrm{reduced}>10$, <16\,mag). The percentage for which additional colour information is available is shown in the table below.}
\label{tab:PTFdata}
\begin{tabular}{lrrrr}
Filter      & \# objects & SDSS &  NOMAD & Pan-STARRS \\
            &            & $ugriz$ &  $JHK$ & $grizy$  \\
\hline
$R$               & $532\,477$    & $43.65\%$ & $97.58\%$ & $98.92\%$ \\
$g^\prime$        & $257\,918$    & $55.45\%$ & $98.69\%$ & $99.26\%$ \\
$R \cap g^\prime$  & $36\,943$     & $64.39\%$ & $96.48\%$ & $98.66\%$ \\
\hline
\end{tabular}
\end{table}

\subsection{Machine Learning Classification}
\label{sec:MLclassifiers}
To cut back on the number of candidates for an initial visual lightcurve inspection, we use supervised machine learning classifiers to make a pre-selection. The idea is that instead of finding EL CVn \added{binaries} by using fixed, pre-defined, user-supplied selection criteria, a sample of known EL CVn \added{binaries} and not--EL CVn objects (a `training set') is provided and a machine learning code (`classifier') decides what the best way is to separate the two groups given the characteristics (called `features', e.g. $g-r$ colour or the lightcurve's root-mean-square value). There are many different types of classifiers, and the behaviour of each classifier can be adjusted by changing so-called hyperparameters. Setting the correct hyperparameters is required to avoid over- or underfitting of the data. For an introduction to machine learning in astronomy, see \citet{astroMLText}, and for a practical guide to machine learning (with \textsc{Python}) see \citet{pythonML}.

Because supervised machine learning classifiers can process huge amounts of data \added{very quickly}, they have become a popular tool to \replaced{handle}{deal} with the large amount of \replaced{lightcurves}{data} produced by survey telescopes. Many different techniques have been tried for lightcurve classification \citep[e.g.][]{2009PhDT.......202D,2013AJ....146..101P,2014A&A...567A.100A,2015ApJ...811...95P,2016ApJ...820..138M,2016MNRAS.456.2260A,2017AJ....153..204S}. In recent years, the Random Forest method \citep{Breiman2001} has become very popular as it typically performs the best and is also easy to interpret \citep[e.g.][]{2011ApJ...733...10R, 2014AJ....148...21M}. 

To find EL CVn binaries we have experimented with three different supervised machine learning classifiers based on combining decision trees: the standard `Random Forest', an `Extra-Trees' classifier \citep{Geurts:2006}, both implemented in the \textsc{Python} package {\sc sklearn} \citep{scikit-learn}, and the more sophisticated `Gradient boosted decision tree' classifier, implemented in {\sc XGBoost} \citep{2016arXiv160302754C}. All three classifiers combine many randomised decision trees, which are a sequence of binary decisions. 

Here we briefly discuss the differences between the methods. Both Random Forest and Extra-Trees combine the prediction of many independent, randomised decision trees. The larger the number of trees the better but \replaced{at the cost of increased}{this increases} computation time. For both methods, each tree is built using only a subset of all features (rule-of-thumb is the square root of the total number of features). Random Forest uses the best possible split of the data given the available features and uses that to separate the different classes. \deleted{The }Extra-Trees differs from Random Forest as it does not use the best split, but a random split. This extra randomization step has the consequence that decision boundaries are more smooth compared to Random Forest. Both methods are relatively simple; they have only a few hyperparameters and are relatively robust against overfitting. 
{\sc XGBoost} also uses many randomised decision trees. But instead of combining many independent trees, \replaced{new trees are created to optimally}{it tries to add trees which} complement the existing trees. This is done by giving samples which were wrongly classified by the previous trees a larger weight when building the next tree. The next tree is, therefore, more likely to classify these objects correctly. The disadvantage of this method is that it is more sensitive to overfitting compared to Random Forest. The {\sc XGBoost} implementation has many hyperparameters which can be set to counteract this, but it can be difficult to determine the best values for these parameters. The advantages of all three methods are that they are insensitive to uninformative features, do not require scaling of the data, and are easy to interpret: they automatically determine the importance of features.

\subsection{EL CVn identification}
Because supervised machine learning \added{algorithms} require a training sample we first need to identify EL CVn binaries in our data. There are no known EL CVn binaries in the PTF magnitude range, so we need to find new ones the old-fashioned way. We do this by selecting a sample of A--type main sequence stars using SDSS colours ($0.8 < u-g_\mathrm{SDSS} < 1.5$ and $-0.5 < g-r_\mathrm{SDSS} < 0.2$) and require that $\text{Stetson-K}>0.6$ (one of the lightcurve statistics, see also \citealt{1996PASP..108..851S}). To limit the sample size and have increased post-facto confidence in the selected objects, we further require that the lightcurve is significantly variable ($\chi^2>40$) and with more than 150 epochs in $R$ and 100 in $g^\prime$. EL CVn binaries are characterized by their $\lesssim 0.1$\,mag, flat-bottomed primary eclipse and slightly shallower secondary eclipse. We, therefore, do a period search using both Analysis-of-Variance and Boxed-Least-Square methods (AoV, \citealt{1989MNRAS.241..153S,2005ApJ...628..411D}; BLS, \citealt{2002A&A...391..369K}, \textsc{vartools} implementation \citealt{2016A&C....17....1H}) on each of the lightcurves and inspect each folded lightcurve for these criteria. In case of doubt the candidate was included in the lightcurve modelling (see Section \ref{sec:LCanalysis}). If the lightcurve fitting showed a `V'-shaped, non-total, eclipse, we rejected it from the sample, as these systems could also be regular MS-MS binary. In other words, we require our systems to be totally eclipsing.

Using this method, we found 6 EL CVn binaries, which we then used as a training set for a Random Forest classifier, combined with a sample of 4000 randomly chosen objects (which we confirmed are not EL CVn binaries). Since the training set is so small, we do not attempt any parameter optimisation, but used the default hyperparameters (500 trees, the square-root of the total number of features as the number of features per tree, and no limits on the tree depth). We applied the classifier to the data (the $\mathrm{PTF}\,R$ and $\mathrm{PTF}\,g^\prime$ lightcurve statistics \replaced{combined with}{and} SDSS colours) and inspected the $\unsim100$ best candidates identified by the classifiers. We added newly found EL CVn candidates to the training sample and repeated the procedure an additional two times. This resulted in the discovery of an additional 11 systems, bringing the total to 17.

Because we required that SDSS colours were available, we only inspected roughly half of the data so far (see Table \ref{tab:PTFdata}). We, therefore, replaced the SDSS colours with the $BVRHJK$ colours from the NOMAD catalogue (Pan-STARRS colours were not yet available at this time). We again checked the best 100 candidates in an iterative way, adding the new EL CVn systems to the training sample. The combined SDSS and NOMAD process resulted in a total of 30 EL CVn binaries. 

With this sample, we trained the three different classifiers (Random Forest, Extra-Trees and {\sc XGBoost}) and determined the best hyperparameters. We use the PTF variability features combined with the Pan-STARRS colours. The goal of our classifier is not to classify all samples correctly (high precision), but instead to rank the candidates according to `EL CVn' likeness. We, therefore, do not optimise the precision of our classifier, but instead we use the `area-under-curve' for the `receiver operating characteristic' (roc-auc). We do this using stratified K-fold cross-validation to calculate the roc-auc score. For more details on classifier metrics (like roc-auc) and model optimisation, see \citet{astroMLText,pythonML}.

For both Random Forest and Extra-Trees we find similar optimal hyperparameters. Using more than 600 trees does not improve performance significantly. The number of features per tree influences the roc-auc score marginally, but there is a clear preference for only using two features per tree. We checked different hyperparameters that limit the depth or complexity of the tree, but we find that the roc-auc score only decreases when the tree depth or complexity is limited using any of the hyperparameters.

For {\sc XGBoost} there are more hyperparameters to tune. We start by optimising the most important three: the number of estimators, the learning rate and the tree depth while setting the other parameters to typical values. After finding the optimal combination of these hyperparameters we continue to optimise the minimum child weight, sub-sampling fraction and the column sub-sample fraction.

After training all classifiers we selected the top 1000 candidates (in both $R$ and $g^\prime$ datasets) from the three classifiers and visually inspected their lightcurves. We found an additional 6 EL CVn binaries, bringing the final number to 36 systems, listed in Table~\ref{tab:geninfo}. 

A quick comparison between the classifiers shows that both the Extra-Trees classifier and the {\sc XGBoost} classifier perform equally well while the Random Forest performs a bit worse. This is confirmed by the ranking of the last six discovered EL CVn binaries that were all further down the list for the Random Forest method. Although Extra-Trees and {\sc XGBoost} performance \replaced{is comparable}{similarly}, tuning the {\sc XGBoost} classifier took significantly more time and effort. Due to the combination of yield versus investment, we deem the Extra-Trees classifier the best (in this particular case).

PTF observed the Kepler field and has thus observed the EL CVn binaries found by Kepler. None of these were recovered by our search and we investigated the reason why. First of all, most Kepler systems feature eclipses much shallower than PTF can detect. The Kepler EL CVn systems with deep enough eclipses to be detected by PTF were not recovered because either the star was saturated in the PTF data, or the object was not observed at a sufficient number of epochs.

\begin{figure*}
\includegraphics{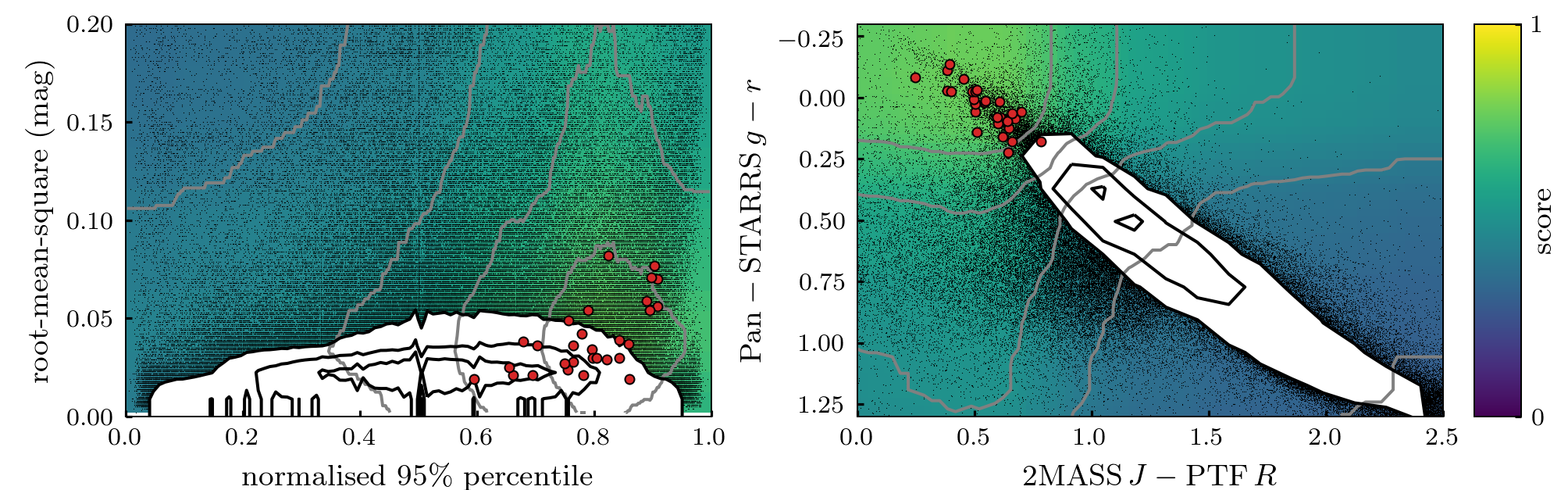}
\caption{The left panel shows the weighted root-mean-squared of the PTF lightcurve versus the normalised 95th percentile of the lightcurve (percentile 95 minus median, divided by 90 percentile range, see Table \ref{tab:featurelist}), the right panel shows 2MASS\,$J$-PTF\,$R$ versus Pan-STARRS g-r colour--colour space. The red dots show all EL CVn binaries, the black contours show all samples in the PTF\,$R$ dataset, with the black contours containing 25, 50, and 75 per cent of the data, samples outsides the contours are indicated with black points. The background colour indicates the `EL CVn'--score by the Extra-Trees classifier with grey lines at every 0.1 score interval. The score is calculated assuming the median values of the EL CVns for all parameters, except the parameters on the x- and y-axis.}
\label{fig:paramspace}
\end{figure*}

\begin{table}
\centering
\caption{Overview of the EL CVn binaries we discovered in the PTF data. In the rest of the paper we will use the PTF name. The ``PTF $R$''-column lists the median magnitude of the light curve in $R$-band.}
\label{tab:geninfo}
\begin{tabular}{llll}
PTF name & IAU name (PTF1 J...) & P (d) & PTF $R$ \\
\hline
PTFS1600y & J004040.23+412521.61 & 1.184 & 13.7 \\
PTFS1600ad & J004300.75+381537.26 & 1.084 & 14.7 \\
PTFS1700do & J005424.06+411126.98 & 3.051 & 15.7 \\
PTFS1600aa & J005659.72+130920.66 & 0.693 & 15.9 \\
PTFS1601p & J011909.91+435907.11 & 1.222 & 15.3 \\
PTFS1501bh & J012814.72+040551.90 & 0.620 & 13.9 \\
PTFS1601q & J013336.92+470600.18 & 1.252 & 16.2 \\
PTFS1601cl & J014839.10+382314.56 & 0.892 & 13.6 \\
PTFS1402de & J021913.15+215921.98 & 0.619 & 15.0 \\
PTFS1607aa & J071207.01+211654.98 & 0.846 & 15.0 \\
PTFS1607v & J075310.42+835154.79 & 0.720 & 15.3 \\
PTFS1607t & J075642.49+162143.99 & 0.876 & 14.2 \\
PTFS1607ab & J075950.03+154319.09 & 0.773 & 14.0 \\
PTFS1608ab & J080425.26+070845.24 & 0.610 & 14.6 \\
PTFS1612al & J121254.27+363341.76 & 0.637 & 15.7 \\
PTFS1512bf & J124154.58+001333.06 & 0.607 & 14.2 \\
PTFS1613s & J133220.59+352847.28 & 1.142 & 14.3 \\
PTFS1613u & J133929.37+455055.64 & 0.564 & 15.3 \\
PTFS1615ag & J150041.84\textminus191417.23 & 0.681 & 14.3 \\
PTFS1615v & J150327.61+460322.78 & 0.559 & 15.9 \\
PTFS1515ay & J150336.10+195842.16 & 0.464 & 14.8 \\
PTFS1615w & J152726.81+120453.54 & 1.441 & 14.9 \\
PTFS1615ao & J152758.90+190751.63 & 0.895 & 15.0 \\
PTFS1615u & J153005.01+202157.06 & 0.778 & 15.8 \\
PTFS1616cr & J162342.13+231456.58 & 0.565 & 14.0 \\
PTFS1617n & J173257.98+403600.93 & 2.337 & 15.3 \\
PTFS1617m & J175433.50+230041.83 & 3.773 & 14.7 \\
PTFS1619l & J191826.08+485302.94 & 1.160 & 13.7 \\
PTFS1521ct & J213318.98+254126.30 & 1.172 & 15.8 \\
PTFS1621ax & J213534.11+233313.86 & 1.018 & 15.0 \\
PTFS1521cm & J214858.33+030417.50 & 0.685 & 15.1 \\
PTFS1622by & J220719.56+085415.66 & 0.749 & 15.8 \\
PTFS1522cc & J225539.41+342137.72 & 0.572 & 14.7 \\
PTFS1622aa & J225652.53+390822.70 & 0.766 & 15.6 \\
PTFS1622bt & J225755.64+310133.67 & 0.688 & 15.1 \\
PTFS1723aj & J231010.08+331249.78 & 1.109 & 14.8 \\
\hline
\end{tabular}
\end{table}

\section{spectroscopy} \label{sec:spectra}
For 19 of our EL CVn systems, we obtained phase-resolved spectroscopy with the Isaac Newton Telescope (INT). We used the Intermediate Dispersion Spectrograph (IDS) with the R632V grating (0.90\,\AA\,$\mathrm{pixel^{-1}}$, 3800--5800\,\AA) for 8 bright nights and the R900V grating for 9 bright nights on 3 separate runs (0.63\,\AA\,$\mathrm{pixel^{-1}}$, 4000--5500\,\AA). Conditions were good with seeing of $\lesssim 1$\,arcsec, except for the last four nights. \replaced{During these nights}{when} the seeing was 2--5\,arcsec and two nights were mostly clouded. An overview of the spectroscopic runs, the set-up and the weather quality is given in Table~\ref{tab:specobs}.

Since the orbital period and phase for all systems is determined very precisely from the photometry (see Section~\ref{sec:results}), we timed the observations such that we observed the systems around orbital phases 0.25 and 0.75. The signal--to--noise per pixel of each spectrum ranges between 40 and 80, sufficient to detect the weaker metal lines in the A/F-star's spectrum. Spectra were taken in pairs and before or after each stellar spectrum a calibration lamp spectrum (CuAr) was obtained to make sure the wavelength calibration was stable. 

The data were reduced using {\sc IRAF}. We used {\sc L.A.Cosmic} \citep{2001PASP..113.1420V} to remove cosmic rays and performed the standard bias and flat calibrations. For the wavelength calibration, we used $\unsim$40 arc lines, which resulted in a typical root-mean-square uncertainty on the wavelength solution of $\lesssim$0.1 pixels (4-6 km\,s$^{-1}$).

\section{Methods and analysis}\label{sec:analysis}

\subsection{Lightcurve}\label{sec:LCanalysis}
By modelling the lightcurves we put strong constraints on the system parameters. To construct a model lightcurve given a set of binary star parameters, we use {\sc lcurve} (by T.R. Marsh and collaborators, see \citealp{2010MNRAS.402.1824C}, see also \citealp{2011MNRAS.410.1113C,2011ApJ...735L..30P}). The {\sc lcurve} code uses grids of points to model the two stars. The shape of the stars in the binary is set by a Roche potential. We assume that the orbit is circular and that the rotation periods of the stars are synchronised to the orbital period. We discuss the validity of the latter assumption in Section \ref{sec:co-rotation}. We calculate the lightcurves assuming the effective wavelength of the PTF filters; 4641\,\AA\ for the $g^\prime$ filter and 6516\,\AA\ for the $R$ filter. In this section (and the rest of the paper), we refer to the A/F--type main sequence as the primary (subscript `1') and the pre-He-WD as the secondary (subscript `2'). 

The free parameters of the model are: the orbital period ($P$) and mid-eclipse time ($t_0$), both in $\mathrm{BMJD_{TDB}}$ (the barycentric Julian date in the terrestrial dynamic time frame, minus 2400000.5), the effective temperatures of both stars ($T_{1,2}$), the scaled radii of both stars ($r_{1,2}=R_{1,2}/a$, where $a$ is the binary separation), the inclination angle ($i$), the mass ratio ($q=M_2/M_1$), an albedo (absorption) for both stars, a linear limb darkening coefficient ($x_{1,2}$), and a gravity darkening coefficient ($y_{1,2}$) in the relation $I\propto g^y$ \citep[where $g$ is the local surface gravity,][]{Zeipel:1924}. Not all these parameters are well constrained by the data and therefore we fix or set an allowed range for some parameters. We constrain the temperature of the primary star ($T_1$) to 6500--10000\,K; the temperature range of A/F--type main-sequence stars. This is needed because with only a lightcurve the temperature ratio is well constrained, but the absolute values of the temperatures of each star are not. We will not use the resulting temperatures of the lightcurve fit, but instead determine the effective temperatures of both stars by modelling the spectral energy distribution (see Section \ref{sec:efftemp}). We fix the limb darkening coefficient of star 2 ($x_2$) to 0.5, since the effect on the lightcurve is minimal. We allow the limb darkening coefficient of \replaced{the A/F-star}{star 1} ($x_1$) to vary between 0.08 to 1.05, the lowest and highest values for stars in the allowed temperature range \citep{2013ApJ...766....3G}. \deleted{We assumed the rotation of both stars to be synchronized to the orbital period. We discuss the validity of this assumption in Section \ref{sec:co-rotation}.} 

To determine the uncertainty on the parameters we use the Markov Chain Monte Carlo (MCMC) method as implemented by \textsc{emcee} \citep{2013PASP..125..306F}. The standard method to determine the uncertainties on the parameters is by using the least-square ($\chi^2$) statistic. However, this assumes that the uncertainty estimates of the data are correct and Gaussian distributed. This is not the case for the PTF lightcurves (as in many observational datasets). Ignoring this problem leads to an underestimate of the uncertainties in the derived parameters, and can in some cases also change the solution. To solve this problem we add additional white noise\footnote{If the noise cannot be treated as white noise, but the noise is correlated (red noise), Gaussian process regression can be used. See for a simple example \citet{george} and an example of this method used to model flickering in a cataclysmic variable by \citet{2017MNRAS.464.1353M}.} to our model (see Section 8 in \citealt{2010arXiv1008.4686H} and for a simple example see \citealt{emcee}). This means that our model has \added{the} white noise amplitude as an extra parameter, which we can simply optimize over, exactly the same as for the lightcurve parameters. 

This method requires the following modification to the standard least-square function: 
\begin{equation}
\label{eq:lc_chi2}
\tilde{\chi}^2 = \sum_n \dfrac{(y_n-m_n (p))^2}{\sigma_n^2 + f^2 m_n(p)^2} + \log(\sigma_n^2 + f^2 m_n(p)^2)
\end{equation}
where $y_n$ is the data, $m_n$ the lightcurve model as a function of the lightcurve parameters $p$, $\sigma_n$ the uncertainties, and $f$ a factor which adds an extra noise source. Note that the first term of the equation is almost the same as in a regular least-square ($\chi^2$) regression, except for the additional noise term [$f^2 m_n(p)^2$]. The first term can be minimised by letting $f$ go to infinity, instead of minimising the difference between data and model [$y_n-m_n (p)$]. Therefore, the second term is needed to penalise models with a large value of $f$. Using this equation, the optimal amount of white noise is added to account for any difference between the data and model. To obtain the best model, we simply minimise $\tilde{\chi}^2$ over the lightcurve parameters $p$ and the parameter $f$, just like regular least-square regression.

For each of the systems, we first find the approximate solution using a simple simplex minimiser of the modified least-square function. We then use \textsc{emcee} to find the best set of parameters of all the available lightcurves for that system. For each filter we use different values for $x_1$, $y_1$, and `absorb', while the rest of the parameters are filter independent. We use 256 parallel MCMC chains (called `walkers') and use at least 2000 generations or more if needed. Any further calculations are done using the last 2560 positions of the walkers.

\subsection{Effective temperature}\label{sec:efftemp}
To determine the temperatures of both components we fit the spectral energy distribution of the target with model spectra, similar as in \citet{2011MNRAS.418.1156M}. We use data from GALEX \citep[far UV \& near UV,][]{2014yCat.2335....0B}, SDSS DR13 \citep[$ugriz$,][]{2016arXiv160802013S}, Pan-STARRS \citep[$grizy$,][]{2016arXiv161205560C}, 2MASS \citep[$HJK$,][]{2006AJ....131.1163S}, and WISE \citep[$W1$ \& $W2$,][]{2010AJ....140.1868W} for each target (where available). We used as model spectra the BaSeL3.1 spectral library \citep{2002A&A...381..524W}. To calculate the flux per band, we convolve the model spectra with each band's response curve.

The overall spectrum is the sum of two model spectra of a given temperature and metallicity, created using bilinear interpolation from the BaSeL library. With only an SED, it is not possible to measure the metallicity of the stars reliably. However, since metallicity and temperature are correlated, we treat \replaced{the metallicity of both stars}{them} as free parameters and marginalise over them in the final result. For the surface gravity, we assume \logg=4 for the A/F-star and \logg=5 for the pre-He-WD. We set the relative contribution to the total light by the ratio between $r_1$ and $r_2$ obtained from the lightcurve. At first, we also used the temperature ratio obtained from the lightcurve, but we learnt that this gave \deleted{wildly} inconsistent predictions for the eclipse depths. This is likely due to the use of blackbody spectra by \textsc{lcurve}. Instead, we directly use the eclipse depth of the primary eclipse \replaced{instead of}{for} the temperature ratio. The final variable is the extinction, set by $E(B-V)$. To calculate the reddening following from the extinction we used the reddening law by \citet{1989ApJ...345..245C} with $R_V=3.1$ (as implemented by \textsc{pysynphot}).

To determine the temperatures of both stars we minimized the function:
\begin{equation}
\label{eq:SEDfit}
\begin{split}
\tilde{\chi}^2 &= \sum_n \dfrac{(y_n-m_n (p))^2}{\sigma_n^2 + f^2 m_n(p)^2} + \log(\sigma_n^2 + f^2 m_n(p)^2) \\ &+ \mathrm{prior}(r_1/r_2,\mathrm{E_{B-V}})
\end{split}
\end{equation}
with $y$ the data, $m$ the model, and $f$ an additional noise factor. We used a value for $E(B-V)$ according to \citet{2011ApJ...737..103S}, with an uncertainty of 0.034 \citep[as in][]{2011MNRAS.418.1156M}. For some added flexibility in our model, we added an extra term of uncertainty to the magnitudes ($f$), similar to the way it was applied in Equation~\ref{eq:lc_chi2}. We again use \textsc{emcee} to determine the best values and uncertainties, similar \replaced{as in}{to} Section~\ref{sec:LCanalysis}.

\subsection{Radial velocity}
To obtain the radial velocity curve of the primary star, we cannot use the Balmer absorption lines in the spectrum because these are present both in the A/F-star and the pre-He-WD. Using these would not \replaced{yield}{give us} reliable results. Instead, we use the metal lines present in the spectra of the A/F--type stars. We cross-correlate the spectra with a template; a high-resolution spectrum of the A5 star HD145689 \citep{2003Msngr.114...10B}. We first interpolate the target spectrum to the (much higher) sampling of HD145689. We then remove the continuum with a low-order polynomial and determine the radial velocity shift using cross-correlation. To estimate the uncertainty on the radial velocity shift, we add random Gaussian noise to the target spectra according to the uncertainty per pixel and measure the radial velocity shift. We repeat this process 11 times and use the standard deviation of the results as the uncertainty. We use the metal lines between 4150--4301\,\AA, 4411--4791\,\AA, and 4941--5400\,\AA\ to get three separate measurements of the radial velocity shift. The radial velocity measurements are corrected to the heliocentric velocity frame with the {\sc rvcorrect} task in {\sc IRAF}.

To determine the radial velocity amplitude, we fit a sinusoidal curve with a fixed value for the period and phase to the measurements. This leaves only the amplitude and systemic velocity as free parameters. We again use a modified least-squares objective function which can also take into account underestimated uncertainties (similar to Equation~\ref{eq:lc_chi2}):
\begin{equation}
\label{eq:rv_chi2}
\tilde{\chi}^2 = \sum_n \dfrac{(y_n-m_n (p))^2}{\sigma_n^2 + f^2} + \log(\sigma_n^2 + f^2)
\end{equation}
with $y_n$ the radial velocity measurements, $\sigma$ the statistical uncertainty on the cross-correlation results and $f$ the extra white noise.
Fitting the data shows (Table \ref{tab:amp}) that $f$ ranges between $7$ and $12\,\mathrm{km\,s^{-1}}$, a factor of 2 higher than the statistical noise $\sigma$. This extra noise is partially due to the uncertainties in the wavelength calibration (4--6\,km\,s$^{-1}$) but does not account for all residual variance. This means that we either underestimate the uncertainties in the cross-correlation procedure (for example by the normalisation of the spectra) or that we underestimate the uncertainties in the calibration process of the spectra. This could be due to instabilities of the optical elements in the IDS/INT combination, which typically changes for each observing run. Since we combine data from four different observing runs, this could result in minor differences in the setup. A potential method to verify this is to check the wavelength of sky emission lines, but these are not available in the spectral range we use. 

\begin{table*}
\centering
\caption{Radial velocity amplitude of the A/F-star, the systemic velocity, the residual variance of the fit, the derived distance, the measured proper motions and the associated stellar population for the 12 EL CVn systems with radial velocity curves. The proper motion is taken from either the UCAC5 catalogue ($^a$, \citealt{2017yCat.1340....0Z}) or
or the Initial Gaia Source List ($^b$, \citealt{2013yCat.1324....0S}).
\label{tab:amp}
}
\begin{tabular}{lrrrrrrl}
ID & $K_1$ & $\gamma$ & $f$ & Distance & $\mu_\alpha$cos$(\delta)$  & $\mu_\delta$ & Pop.\\
   & $(\mathrm{km\,s^{-1}})$ & $(\mathrm{km\,s^{-1}})$ & $(\mathrm{km\,s^{-1}})$ & (pc) & $(\mathrm{mas\,yr^{-1}})$ & $(\mathrm{mas\,yr^{-1}})$  & \\ 
\hline
PTFS1600y & $22.8 \pm 0.9$ & $-88.7 \pm 0.9$ & 6.9  & $2340\pm\hphantom{0}70$ & $-7.3\pm2.0^a$ & $-4.2\pm1.3^a$ & Thin \\
PTFS1600ad & $29.7 \pm 1.4$ & $-23.3 \pm 1.2$ & 7.2   & $3770\pm180$ & $1.1\pm1.5^a$ & $-1.8\pm1.5^a$ & Thin  \\
PTFS1601p & $18.4 \pm 2.1$ & $-45.9 \pm 1.8$ & 12.4  & $4960\pm500$ & $1.4\pm4.6^a$ & $-7.6\pm4.2^a$ & Thin/Thick \\
PTFS1501bh & $24.0 \pm 1.5$ & $16.6 \pm 1.2$ & 8.2  & $1280\pm\hphantom{0}70$ & $10.1\pm 1.5^a$ & $-4.9\pm1.5^a$ & Thin \\
PTFS1601cl & $35.2 \pm 2.1$ & $-14.4 \pm 1.2$ & 9.1  & $2890\pm\hphantom{0}90$ & $2.0\pm1.3^a$ & $2.2\pm1.3^a$ & Thin \\
PTFS1607t & $26.7 \pm 1.7$ & $31.0 \pm 1.2$ & 5.5  & $2160\pm\hphantom{0}70$ & $-2.1\pm0.3^a$ & $1.6\pm1.6^a$ & Thin \\
PTFS1607ab & $32.7 \pm 1.3$ & $-37.6 \pm 1.1$ & 6.3  & $1810\pm\hphantom{0}70$ & $-3.1\pm2.6^b$ & $-8.2\pm3.1^b$ & Thin \\
PTFS1512bf & $31.4 \pm 1.9$ & $70.3 \pm 1.5$ & 11.8 & $1820\pm\hphantom{0}50$ & $-19.2\pm6.4^a$ & $4.7\pm5.7^a$ & Thick \\
PTFS1617n & $17.7 \pm 1.7$ & $-198.7 \pm 1.9$ & 7.2  & $5700\pm380$ & $-1.8\pm2.6^a$ & $-2.9\pm2.7^a$& Thick/Halo  \\
PTFS1617m & $13.1 \pm 2.4$ & $-40.4 \pm 2.1$ & 10.9  & $4060\pm180$ & $-0.9\pm1.5^a$ & $-9.2\pm1.7^a$ & Thin/Thick \\
PTFS1619l & $22.7 \pm 1.3$ & $-12.5 \pm 1.0$ & 5.8   & $2040\pm140$ & $-1.9\pm1.6^a$ & $-2.0\pm1.6^a$ & Thin \\
PTFS1521cm & $34.7 \pm 2.0$ & $-6.5 \pm 1.9$ & 7.9   & $2870\pm110$ & $9.6\pm2.8^a$ & $-7.8\pm2.7^a$ & Thick \\
\hline
\end{tabular}
\end{table*}

\subsection{Galactic kinematics}\label{sec:galkin}
For the 12 systems for which we have obtained a radial velocity measurement, we calculate their Galactic location and velocity. We determine the distance to the systems by using the $K$-band magnitude and absolute radius, combined with the $K$-band surface brightness calibration by \citet{2008A&A...491..855K}. The proper motions of the systems are taken from the UCAC5 catalogue \citep{2017yCat.1340....0Z} or the Initial Gaia Source List \citep{2013yCat.1324....0S}. Combined with right ascension and declination, we calculate velocity in the direction of the Galactic Centre ($V_\rho$) and the Galactic rotation direction ($V_\phi$), the Galactic orbital eccentricity ($e$), and the angular momentum in the Galactic $z$ direction ($J_z$). The Galactic radial velocity $V_\rho$ is negative towards the Galactic centre, while stars that are revolving on retrograde orbits around the Galactic Centre have negative $V_\phi$. Stars on retrograde orbits have positive $J_z$. Thin disk stars generally have very low eccentricities $e$. Population membership can be derived from the position in the $V_\rho$ - $V_\phi$ diagram and the $J_z$ - $e$ diagram \citep{2006A&A...447..173P}. 

\begin{figure*}
\includegraphics[width=0.48\textwidth]{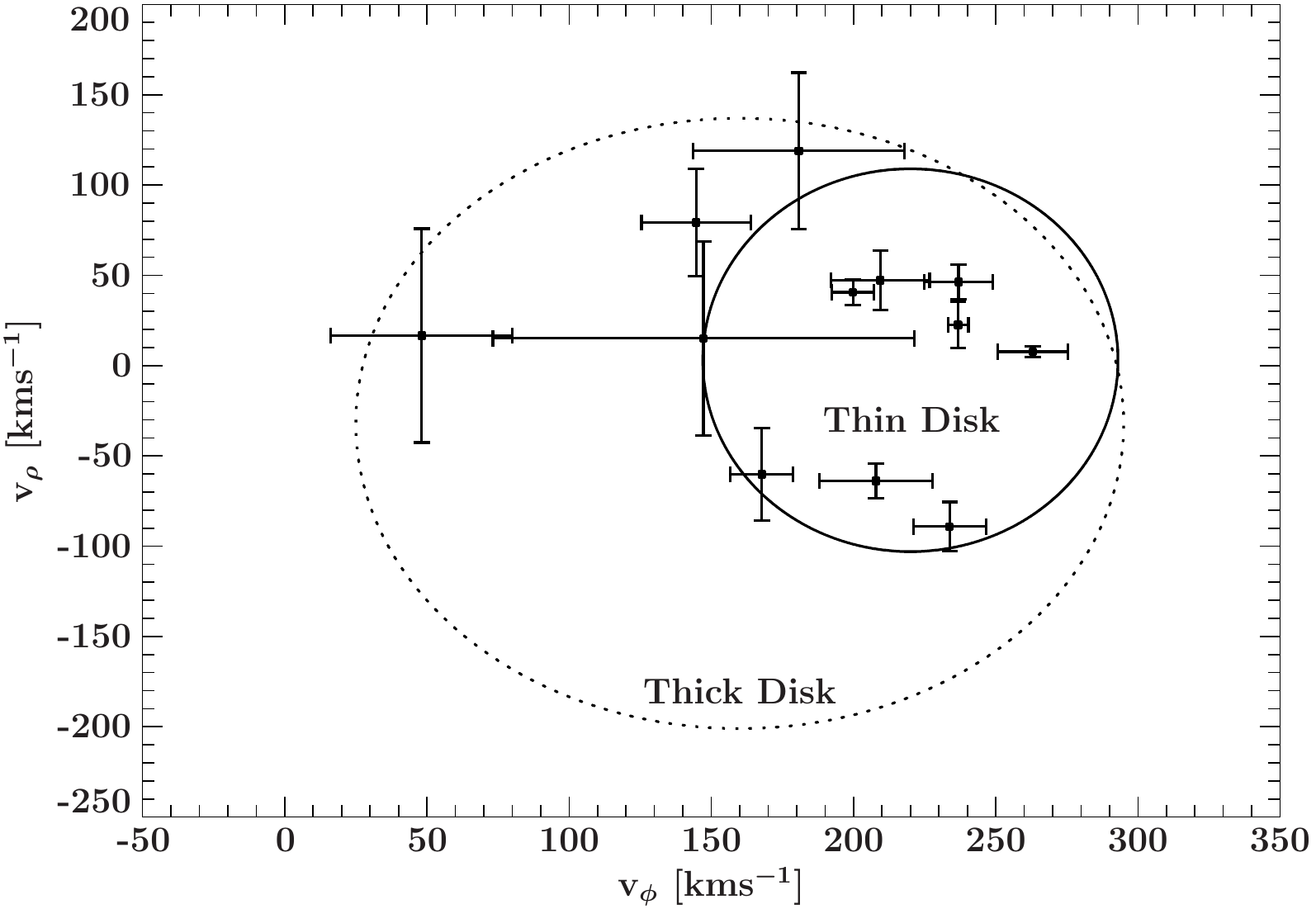}\hfill
\includegraphics[width=0.48\textwidth]{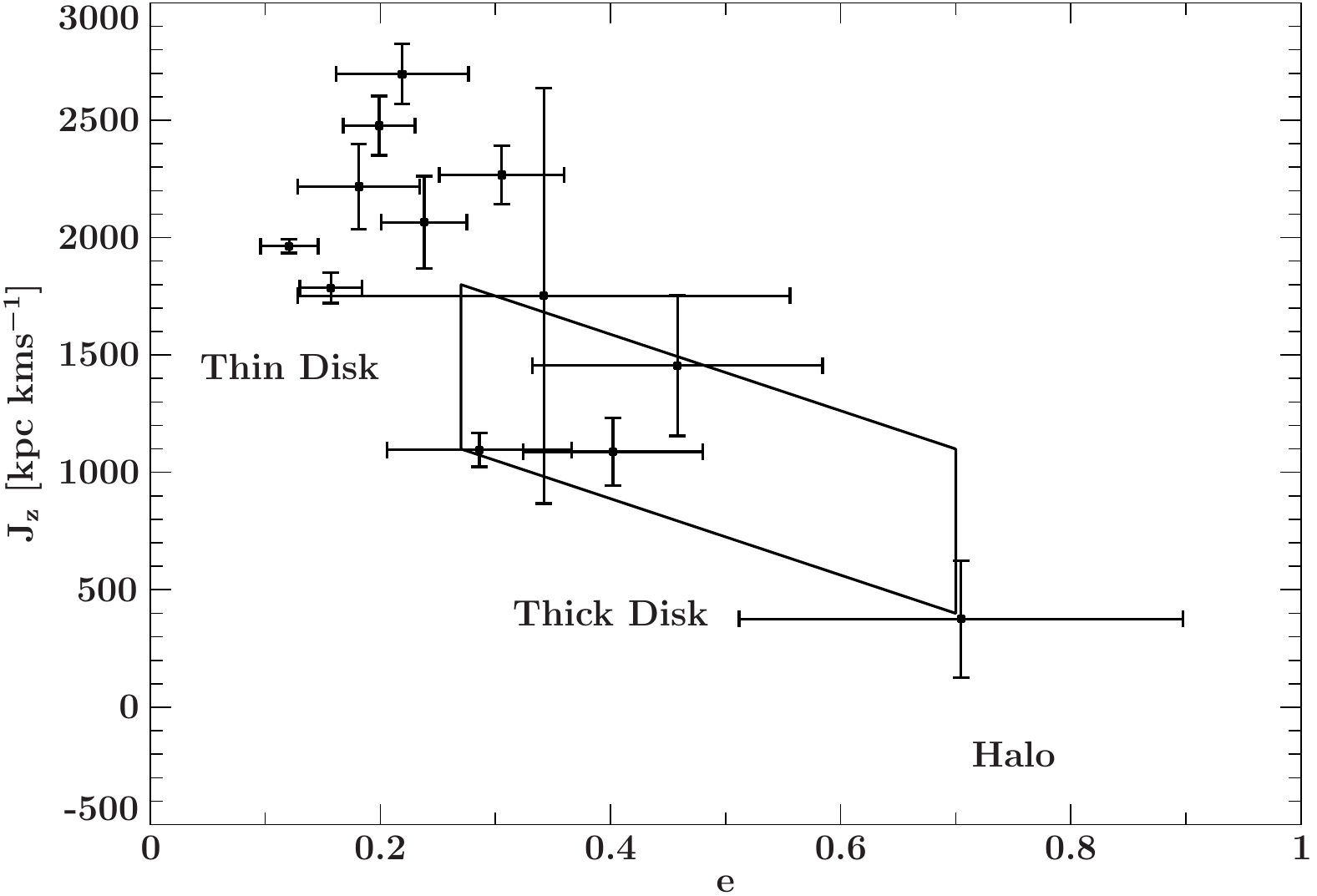}\\
\vspace{0.2cm}
\caption{$V_\phi$--$V_\rho$ (left) and $e$--$J_{z}$ diagrams (right). The solid and dotted ellipses render the 3$\sigma$ thin and thick disk contours in the $V_\phi$--$V_\rho$ diagram, while the solid box in the $e$--$J_{z}$ marks the thick disk region as specified by \citet{2006A&A...447..173P}.} 
\label{fig:galkin}
\end{figure*}

\subsection{Masses and radii}\label{sec:calcMR}
\begin{figure}
\includegraphics{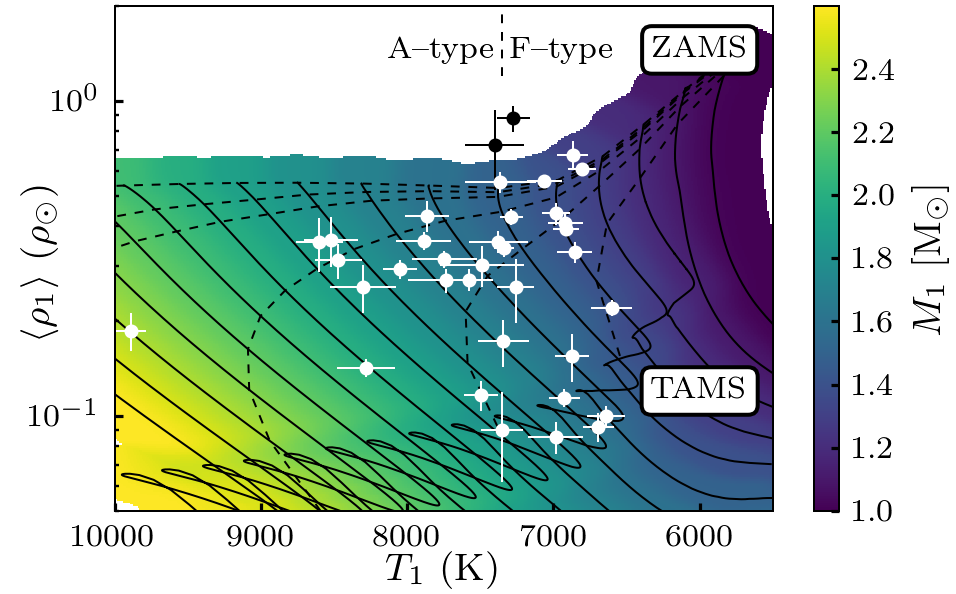}
\caption{The temperature versus the average density of the main-sequence stars of the binary system, indicated with a black or white dot. The black lines show main-sequence evolution tracks (solar metallicity) from \citet{2017ApJ...838..161S} between 1.0 and 2.5\,\Msun\ with 0.1\,\Msun\ intervals. The colour map shows the mass of the star according to the interpolation method by \citet{2012ApJ...748..115B}. The dashed lines are isochrones of 0, 0.1, 0.2, 0.5, 1, and 2\,Gyr since the start of the main sequence.}
\label{fig:density}
\end{figure}

To fully solve for the elements of the binary system, we need to combine the information from the lightcurve fit with an additional piece of information to set the scale of the system. This is typically done by measuring the radial velocity amplitude of both stars. We only have the radial velocity amplitude of one of the stars in the binary. In principle, we can combine this with the mass-ratio $q$, but the uncertainties on the mass-ratio derived from the lightcurve fitting are high, and the uncertainty on the masses scale with a high power of $q$ (for low $q$: $M_1\propto K_1^2q^{-3}$, $M_2\propto K_1^3q^{-2}$), and are therefore not constraining. 

To circumvent this problem, we make use of the assumption that the primary star is a main-sequence star. Using only the lightcurve parameters, we can calculate the average density of the main-sequence component:
\begin{equation}
\label{eq:rho}
\langle\rho\rangle = \frac{3\upi}{G P^2 r_1^3 (1+q)}
\end{equation}
To propagate the uncertainties correctly, we calculate the stellar density for each point in the MCMC chain and assign a random temperature according to our measurement of the SED. With the average density and temperature of the main-sequence star, we can use stellar models to determine its mass. We use the Yale-Potsdam stellar models \citep{2017ApJ...838..161S} and follow the same procedure as in \citet{2012ApJ...748..115B} to make a continues mapping of the mass in $T-\langle \rho \rangle$ space. We convolve each track with a Gaussian probability function with a standard deviation of $200$K in temperature and 0.1 dex in density. For each point in the temperature-density grid, we assign the mass with the highest probability. We use this mapping to calculate the primary mass for the posterior distribution of the lightcurve fits (see Section \ref{sec:LCanalysis}). As can be seen in Fig.~\ref{fig:density}, most but not all measurements agree with the models. Two systems, PTFS1612al and PTFS1615u, have slightly higher densities than would be the case for a solar metallicity composition for zero-age main sequence models. For these two systems, we extrapolate the models to determine the mass.

With the mass of the primary ($M_1$) combined with $q$, $i$ and $P$, we calculate the semi-major axis ($a$) using Kepler's law; \begin{equation}
\label{eq:sma}
a^3 = GM_1\,(1+q)\ \left(\frac{2\upi}{P}\right)^2
\end{equation}
Note that in both equations the mass ratio is present in the form of $1+q$, and since the mass ratio is small ($q\sim0.1$), the high uncertainty on $q$ only mildly affects the accuracy on $a$ and $M_1$. However, the uncertainty on the pre-He-WDs mass ($M_2=q M_1$) is proportional to the uncertainty on $q$, which means that the uncertainty on $M_2$ is too high to be constraining. 

This can be solved by including the measured radial velocity ($K_1$) in our calculation, which is available for 12 systems. We use an iterative approach to find the optimal solution as in \citet{2015ApJ...803...82R}, again for each sample from the lightcurve fit posterior distribution. This involves calculating $M_1$, $q$ and $\langle\rho\rangle$ until the solution converges, which it does after 2 iterations.

\section{Results}\label{sec:results}
\begin{table*}
\caption{System parameters of all EL CVn systems and the \replaced{uncertainty (standard deviation)}{standard deviation} on the parameters. Systems for which a radial velocity measurement is used to calculate the parameters is indicated with the `RV' superscript. This mainly affects the reliability and systematics on the mass and surface gravity of the pre-He-WD ($M_2$ and $\log g_2$).}
\label{tab:syspar}
\begin{tabular}{@{}l *{10}{r} }
Name & $P$\,(d) & $i$\,(\degr) & $M_1$\,(\Msun) & $M_2$\,(\Msun) & $R_1$\,(\Rsun) & $R_2$\,(\Rsun) & $T_1$\,(K) & $T_2$\,(K) & $\log g_1$ & $\log g_2$ \\
\hline
\noalign{\smallskip}
1600y$^\mathrm{RV}$ & $ 1.1838920$  & $      84.5$  & $      1.62$  & $      0.17$  & $      2.41$  & $      0.46$  & $6930$  & $8900$  & $      3.88$  & $      4.33$  \\
 & $ 0.0000008$  & $       2.7$  & $      0.04$  & $      0.01$  & $      0.07$  & $      0.02$  & $100$  & $110$  & $      0.02$  & $      0.03$  \\
\noalign{\smallskip}
1600ad$^\mathrm{RV}$ & $ 1.0840448$  & $      86.5$  & $      1.76$  & $      0.23$  & $      1.83$  & $      0.35$  & $8050$  & $10490$  & $      4.16$  & $      4.72$  \\
 & $ 0.0000010$  & $       2.2$  & $      0.04$  & $      0.01$  & $      0.05$  & $      0.02$  & $120$  & $200$  & $      0.02$  & $      0.04$  \\
\noalign{\smallskip}
1700do & $ 3.0507582$  & $      87.4$  & $      2.40$  & $      0.81$  & $      2.34$  & $      0.33$  & $9890$  & $17100$  & $      4.08$  & $      5.31$  \\
 & $ 0.0000278$  & $       1.8$  & $      0.06$  & $      0.25$  & $      0.13$  & $      0.03$  & $90$  & $1400$  & $      0.04$  & $      0.15$  \\
\noalign{\smallskip}
1600aa & $ 0.6934558$  & $      78.7$  & $      1.67$  & $      0.50$  & $      1.67$  & $      0.55$  & $7880$  & $9300$  & $      4.21$  & $      4.67$  \\
 & $ 0.0000006$  & $       0.9$  & $      0.05$  & $      0.09$  & $      0.04$  & $      0.02$  & $190$  & $400$  & $      0.02$  & $      0.09$  \\
\noalign{\smallskip}
1601p$^\mathrm{RV}$ & $ 1.2215885$  & $      83.8$  & $      1.82$  & $      0.14$  & $      1.65$  & $      0.34$  & $8600$  & $11700$  & $      4.26$  & $      4.54$  \\
 & $ 0.0000051$  & $       3.2$  & $      0.06$  & $      0.02$  & $      0.14$  & $      0.04$  & $160$  & $500$  & $      0.06$  & $      0.10$  \\
\noalign{\smallskip}
1501bh$^\mathrm{RV}$ & $ 0.6204144$  & $      78.4$  & $      1.30$  & $      0.12$  & $      1.23$  & $      0.20$  & $6870$  & $11100$  & $      4.38$  & $      4.91$  \\
 & $ 0.0000005$  & $       1.9$  & $      0.04$  & $      0.01$  & $      0.07$  & $      0.01$  & $110$  & $400$  & $      0.04$  & $      0.06$  \\
\noalign{\smallskip}
1601q & $ 1.2515058$  & $      80.5$  & $      1.85$  & $      0.30$  & $      1.93$  & $      0.46$  & $8300$  & $10700$  & $      4.13$  & $      4.58$  \\
 & $ 0.0000051$  & $       2.9$  & $      0.08$  & $      0.18$  & $      0.15$  & $      0.04$  & $200$  & $700$  & $      0.05$  & $      0.40$  \\
\noalign{\smallskip}
1601cl$^\mathrm{RV}$ & $ 0.8917354$  & $      82.9$  & $      2.02$  & $      0.28$  & $      2.44$  & $      0.52$  & $8290$  & $10100$  & $      3.97$  & $      4.45$  \\
 & $ 0.0000005$  & $       2.9$  & $      0.06$  & $      0.01$  & $      0.07$  & $      0.02$  & $200$  & $300$  & $      0.02$  & $      0.03$  \\
\noalign{\smallskip}
1402de & $ 0.6189694$  & $      87.0$  & $      1.61$  & $      0.36$  & $      1.56$  & $      0.45$  & $7860$  & $9300$  & $      4.27$  & $      4.69$  \\
 & $ 0.0000011$  & $       2.5$  & $      0.04$  & $      0.13$  & $      0.07$  & $      0.02$  & $150$  & $300$  & $      0.03$  & $      0.23$  \\
\noalign{\smallskip}
1607aa & $ 0.8463120$  & $      84.6$  & $      1.85$  & $      0.30$  & $      1.81$  & $      0.38$  & $8470$  & $10300$  & $      4.19$  & $      4.76$  \\
 & $ 0.0000016$  & $       3.4$  & $      0.05$  & $      0.07$  & $      0.08$  & $      0.02$  & $160$  & $300$  & $      0.03$  & $      0.13$  \\
\noalign{\smallskip}
1607v & $ 0.7198356$  & $      82.6$  & $      1.58$  & $      0.20$  & $      1.83$  & $      0.16$  & $7260$  & $10900$  & $      4.11$  & $      5.32$  \\
 & $ 0.0000020$  & $       5.8$  & $      0.06$  & $      0.05$  & $      0.16$  & $      0.03$  & $120$  & $500$  & $      0.06$  & $      0.24$  \\
\noalign{\smallskip}
1607t$^\mathrm{RV}$ & $ 0.8759507$  & $      76.6$  & $      1.40$  & $      0.16$  & $      1.87$  & $      0.38$  & $6600$  & $8600$  & $      4.04$  & $      4.48$  \\
 & $ 0.0000004$  & $       1.0$  & $      0.05$  & $      0.01$  & $      0.05$  & $      0.01$  & $140$  & $200$  & $      0.02$  & $      0.03$  \\
\noalign{\smallskip}
1607ab$^\mathrm{RV}$ & $ 0.7730986$  & $      83.8$  & $      1.40$  & $      0.19$  & $      1.45$  & $      0.32$  & $6980$  & $8810$  & $      4.26$  & $      4.71$  \\
 & $ 0.0000002$  & $       2.3$  & $      0.03$  & $      0.01$  & $      0.05$  & $      0.01$  & $100$  & $80$  & $      0.03$  & $      0.04$  \\
\noalign{\smallskip}
1608ab & $ 0.6101718$  & $      86.8$  & $      1.50$  & $      0.11$  & $      1.39$  & $      0.52$  & $7400$  & $7900$  & $      4.32$  & $      4.04$  \\
 & $ 0.0000014$  & $       1.9$  & $      0.06$  & $      0.10$  & $      0.04$  & $      0.02$  & $200$  & $400$  & $      0.02$  & $      0.24$  \\
\noalign{\smallskip}
1612al & $ 0.6369260$  & $      86.8$  & $      1.38$  & $      0.16$  & $      1.16$  & $      0.37$  & $7280$  & $10300$  & $      4.45$  & $      4.50$  \\
 & $ 0.0000006$  & $       2.0$  & $      0.08$  & $      0.09$  & $      0.05$  & $      0.02$  & $110$  & $300$  & $      0.03$  & $      0.22$  \\
\noalign{\smallskip}
1512bf$^\mathrm{RV}$ & $ 0.6074343$  & $      87.2$  & $      1.39$  & $      0.17$  & $      1.53$  & $      0.32$  & $6910$  & $9740$  & $      4.21$  & $      4.65$  \\
 & $ 0.0000002$  & $       1.9$  & $      0.02$  & $      0.01$  & $      0.02$  & $      0.01$  & $90$  & $180$  & $      0.01$  & $      0.04$  \\
\noalign{\smallskip}
1613s & $ 1.1420695$  & $      76.2$  & $      1.83$  & $      0.17$  & $      2.72$  & $      0.25$  & $7350$  & $13700$  & $      3.83$  & $      4.88$  \\
 & $ 0.0000024$  & $       6.0$  & $      0.08$  & $      0.05$  & $      0.30$  & $      0.05$  & $140$  & $800$  & $      0.08$  & $      0.28$  \\
\noalign{\smallskip}
1613u & $ 0.5644902$  & $      81.6$  & $      1.52$  & $      0.27$  & $      1.65$  & $      0.37$  & $7340$  & $9690$  & $      4.19$  & $      4.73$  \\
 & $ 0.0000003$  & $       2.7$  & $      0.02$  & $      0.04$  & $      0.04$  & $      0.01$  & $70$  & $160$  & $      0.02$  & $      0.08$  \\
\noalign{\smallskip}
1615ag & $ 0.6806897$  & $      85.7$  & $      1.52$  & $      0.27$  & $      1.63$  & $      0.32$  & $7370$  & $10200$  & $      4.20$  & $      4.87$  \\
 & $ 0.0000046$  & $       3.3$  & $      0.05$  & $      0.07$  & $      0.05$  & $      0.02$  & $200$  & $400$  & $      0.02$  & $      0.13$  \\
\noalign{\smallskip}
1615v & $ 0.5594054$  & $      73.7$  & $      1.39$  & $      0.13$  & $      1.50$  & $      0.33$  & $6920$  & $9400$  & $      4.23$  & $      4.54$  \\
 & $ 0.0000003$  & $       1.4$  & $      0.03$  & $      0.03$  & $      0.04$  & $      0.01$  & $120$  & $300$  & $      0.02$  & $      0.17$  \\
\noalign{\smallskip}
1515ay & $ 0.4642873$  & $      89.0$  & $      1.33$  & $      0.15$  & $      1.30$  & $      0.46$  & $6800$  & $7930$  & $      4.33$  & $      4.27$  \\
 & $ 0.0000001$  & $       1.1$  & $      0.03$  & $      0.04$  & $      0.02$  & $      0.01$  & $100$  & $150$  & $      0.01$  & $      0.10$  \\
\noalign{\smallskip}
1615w & $ 1.4407151$  & $      77.7$  & $      1.61$  & $      0.24$  & $      2.59$  & $      0.40$  & $6690$  & $10300$  & $      3.82$  & $      4.63$  \\
 & $ 0.0000024$  & $       2.7$  & $      0.05$  & $      0.05$  & $      0.14$  & $      0.03$  & $110$  & $200$  & $      0.04$  & $      0.14$  \\
\noalign{\smallskip}
1615ao & $ 0.8954515$  & $      77.6$  & $      1.64$  & $      0.41$  & $      1.82$  & $      0.64$  & $7580$  & $8700$  & $      4.13$  & $      4.43$  \\
 & $ 0.0000007$  & $       0.8$  & $      0.05$  & $      0.12$  & $      0.05$  & $      0.03$  & $170$  & $160$  & $      0.02$  & $      0.15$  \\
\noalign{\smallskip}
1615u & $ 0.7777349$  & $      82.4$  & $      1.50$  & $      0.24$  & $      1.27$  & $      0.16$  & $7400$  & $12200$  & $      4.40$  & $      5.41$  \\
 & $ 0.0000011$  & $       4.2$  & $      0.09$  & $      0.10$  & $      0.14$  & $      0.02$  & $200$  & $600$  & $      0.08$  & $      0.27$  \\
\noalign{\smallskip}
1616cr & $ 0.5649690$  & $      82.5$  & $      1.40$  & $      0.07$  & $      1.36$  & $      0.46$  & $7060$  & $8000$  & $      4.32$  & $      3.93$  \\
 & $ 0.0000002$  & $       0.9$  & $      0.03$  & $      0.02$  & $      0.02$  & $      0.01$  & $120$  & $170$  & $      0.01$  & $      0.07$  \\
\hline
\end{tabular}
\end{table*}

\begin{table*}
\contcaption{}
\begin{tabular}{@{}l *{10}{r} }
Name & $P$\,(d) & $i$\,(\degr) & $M_1$\,(\Msun) & $M_2$\,(\Msun) & $R_1$\,(\Rsun) & $R_2$\,(\Rsun) & $T_1$\,(K) & $T_2$\,(K) & $\log g_1$ & $\log g_2$ \\
\hline
1617n$^\mathrm{RV}$ & $ 2.3367776$  & $      87.3$  & $      1.80$  & $      0.18$  & $      2.41$  & $      0.38$  & $7500$  & $11600$  & $      3.93$  & $      4.55$  \\
 & $ 0.0000052$  & $       2.1$  & $      0.04$  & $      0.02$  & $      0.12$  & $      0.03$  & $110$  & $400$  & $      0.04$  & $      0.07$  \\
\noalign{\smallskip}
1617m$^\mathrm{RV}$ & $ 3.7728999$  & $      87.8$  & $      1.68$  & $      0.14$  & $      2.57$  & $      0.69$  & $6990$  & $9320$  & $      3.84$  & $      3.89$  \\
 & $ 0.0000083$  & $       1.5$  & $      0.06$  & $      0.03$  & $      0.08$  & $      0.03$  & $190$  & $190$  & $      0.02$  & $      0.10$  \\
\noalign{\smallskip}
1619l$^\mathrm{RV}$ & $ 1.1599993$  & $      83.2$  & $      1.56$  & $      0.17$  & $      2.13$  & $      0.34$  & $6870$  & $9200$  & $      3.97$  & $      4.60$  \\
 & $ 0.0000017$  & $       4.1$  & $      0.05$  & $      0.01$  & $      0.14$  & $      0.04$  & $120$  & $150$  & $      0.05$  & $      0.09$  \\
\noalign{\smallskip}
1521ct & $ 1.1724964$  & $      83.0$  & $      1.82$  & $      0.36$  & $      1.72$  & $      0.56$  & $8520$  & $9800$  & $      4.23$  & $      4.50$  \\
 & $ 0.0000014$  & $       1.4$  & $      0.06$  & $      0.29$  & $      0.11$  & $      0.04$  & $190$  & $300$  & $      0.04$  & $      0.28$  \\
\noalign{\smallskip}
1621ax & $ 1.0181522$  & $      84.0$  & $      1.69$  & $      0.30$  & $      2.14$  & $      0.17$  & $7350$  & $11800$  & $      4.00$  & $      5.48$  \\
 & $ 0.0000045$  & $       4.6$  & $      0.06$  & $      0.07$  & $      0.14$  & $      0.03$  & $170$  & $700$  & $      0.04$  & $      0.22$  \\
\noalign{\smallskip}
1521cm$^\mathrm{RV}$ & $ 0.6854774$  & $      80.0$  & $      1.49$  & $      0.21$  & $      1.49$  & $      0.43$  & $7290$  & $9240$  & $      4.27$  & $      4.49$  \\
 & $ 0.0000002$  & $       1.0$  & $      0.02$  & $      0.01$  & $      0.03$  & $      0.01$  & $70$  & $90$  & $      0.02$  & $      0.03$  \\
\noalign{\smallskip}
1622by & $ 0.7486683$  & $      85.8$  & $      1.69$  & $      0.31$  & $      1.84$  & $      0.33$  & $7700$  & $11100$  & $      4.13$  & $      4.88$  \\
 & $ 0.0000016$  & $       3.4$  & $      0.07$  & $      0.07$  & $      0.07$  & $      0.02$  & $300$  & $1400$  & $      0.03$  & $      0.13$  \\
\noalign{\smallskip}
1522cc & $ 0.5717853$  & $      81.2$  & $      1.40$  & $      0.26$  & $      1.62$  & $      0.27$  & $6860$  & $9570$  & $      4.17$  & $      4.99$  \\
 & $ 0.0000003$  & $       3.1$  & $      0.04$  & $      0.04$  & $      0.05$  & $      0.01$  & $120$  & $190$  & $      0.02$  & $      0.08$  \\
\noalign{\smallskip}
1622aa & $ 0.7661291$  & $      84.7$  & $      1.60$  & $      0.16$  & $      1.74$  & $      0.26$  & $7500$  & $10900$  & $      4.16$  & $      4.85$  \\
 & $ 0.0000038$  & $       4.0$  & $      0.08$  & $      0.05$  & $      0.10$  & $      0.03$  & $300$  & $1300$  & $      0.04$  & $      0.19$  \\
\noalign{\smallskip}
1622bt & $ 0.6884160$  & $      79.2$  & $      1.65$  & $      0.29$  & $      1.74$  & $      0.29$  & $7700$  & $12200$  & $      4.18$  & $      4.97$  \\
 & $ 0.0000004$  & $       2.0$  & $      0.06$  & $      0.04$  & $      0.05$  & $      0.01$  & $200$  & $1000$  & $      0.02$  & $      0.08$  \\
\noalign{\smallskip}
1723aj & $ 1.1088064$  & $      85.6$  & $      1.57$  & $      0.18$  & $      2.51$  & $      0.23$  & $6640$  & $11000$  & $      3.84$  & $      4.98$  \\
 & $ 0.0000009$  & $       3.3$  & $      0.05$  & $      0.03$  & $      0.07$  & $      0.02$  & $130$  & $400$  & $      0.02$  & $      0.11$  \\
\noalign{\smallskip}
\hline
\end{tabular}
\end{table*}

For the 36 EL CVn binaries in the PTF data, we fit the lightcurves with a binary star model, see Fig.~\ref{fig:LCs} and Table~\ref{tab:lcstats} in the Appendix. The best model fits to the lightcurves all show a flat-bottomed primary eclipse and a round-bottom secondary eclipse. The orbital period of the binary, the radii of both stars, and the orbital inclination are typically well constrained, but the uncertainty on the mass-ratios of the systems is typically $\gtrsim10$\,per cent. The extra noise term in the fit for the lightcurves is typically $\lesssim1$\,per cent. This is consistent with the expected uncertainty in the absolute photometric calibration which is not part of the error bars of the lightcurves. The orbital periods of the binaries range from 0.46\,d to 3.8\,d, with inclinations between 74--90 degrees. The radii of the primary stars divided by the semi-major axis ($r_1$) are typically 0.2--0.5, and the primary stars fill about 0.4--0.9 of their Roche lobe. The average density derived from the lightcurve is typically between 10--70 per cent of Solar density, consistent with A/F--type main sequence stars. The mass-ratio as determined from the lightcurves are typically between 0.08--0.2, but there are outliers to larger ratios. However, the uncertainties on the outliers are high. 
\replaced{The mass-ratio determined from the amplitude of the inter-eclipse variability, which is in some cases not significant (e.g. 1700do) and explains the high uncertainty on the value for the mass-ratio is some cases.}{The accuracy depends on the mass-ratio which is determined from the amplitude of the inter-eclipse variability, which is in some cases not significant (e.g. 1700do) and explains the high uncertainty on the mass-ratio is some cases.} From the lightcurve we determined the temperature ratio of the two stars, assuming blackbody spectral energy distributions is typically 0.5--0.95. 

The results of the SED fitting \deleted{to determine the temperatures} are shown in Table~\ref{tab:syspar} and Fig.~\ref{fig:SEDcurves}. The temperatures of the A/F-stars in the EL CVn systems range between 6600--10000\,K, consistent with temperatures for A/F--type main sequence stars (F-type: 6000--7350\,K, A-type: 7350--10000\,K, \citealt{2013ApJS..208....9P}). The temperature of the pre-He-WDs range from 7900 to 17000\,K. In all systems, the pre-He-WDs are hotter than the A/F-star companion. The uncertainty on the A/F-star's temperature is typically 100--200\,K. The temperature of the pre-He-WD is less well constrained (100--1400\,K), because it depends on the availability of UV data and on how accurately the eclipse depth can be measured from the lightcurves. The RMS scatter between the data and model is typically $\lesssim5$\,per cent, with a few outliers to 10\,per cent (see Table~\ref{tab:SEDfit}). This residual scatter can be due to calibration differences between telescopes, but also because the observations are taken at a random phase. An observation taken in-eclipse results in a $\unsim10$\,per cent lower flux as out of eclipse.

For 12 of the EL CVn systems we obtain usable radial velocity curves and measure the radial velocity amplitude, see Fig.~\ref{fig:RVcurves}. The remaining 7 systems were \added{not enough measurements were obtained or they were} observed at unfavourable orbital phases, precluding an accurate radial velocity measurement. \replaced{However,}{Although the uncertainty on the individual measurements can be large,} all radial velocity amplitudes are low, in the range of 20--40\,$\mathrm{km\,s^{-1}}$. This confirms that the secondary stars in these binaries are indeed low mass stars. 

Using all information available, we determine the stellar parameters of the stars in the EL CVn systems, summarised in Table~\ref{tab:syspar}. The masses of the A/F-type star range between 1.3 and 2.4\,\Msun. The radii of these stars (1.15--2.7\,\Rsun) are consistent with these stars being regular main-sequence stars. 

The radii of the pre-He-WDs range between 0.17--0.65\,\Rsun. To calculate the mass of the pre-He-WD ($M_2$), we include the measured radial velocity amplitude if available, which `replaces' the uncertain mass-ratio measurement from the lightcurve. For most of the systems we do not have a radial velocity amplitude measurement, so we do depend on the mass-ratio to determine the mass of the pre-He-WDs, which range between 0.12 and 0.5\,\Msun. As discussed in Section \ref{sec:LCanalysis} and \ref{sec:calcMR}, the mass determination of the secondary using only the mass-ratio is very uncertain because of the high uncertainty on the mass-ratio. If we limit ourselves only to systems for which we have a radial velocity amplitude, the mass range is $0.12$--$0.28$\,\Msun, significantly smaller. 

For the sample for which we have radial velocity curves, we determine the motion in the Galactic plane and derive their population membership as described in Sec.\,\ref{sec:galkin} and shown in Table\ \ref{tab:amp}. Fig.~\ref{fig:galkin} shows that more than half of the systems are part of the thin disk population. A few are part of the thick disk, and PTFS1617n could also be a halo object.

\section{Discussion}\label{sec:discussion}

\subsection{Co-rotation}\label{sec:co-rotation}

In the lightcurve modelling (Section \ref{sec:LCanalysis}), we assume that both stars are synchronised with the orbit. Previous studies of EL CVn binaries have made the same assumption, but all authors acknowledge that it might not be correct, since mass-accretion can spin up the A/F--star significantly (see Section \ref{sec:evolution}). \citet{2010ApJ...715...51V} extensively discuss how all parameter estimates are affected by incorrectly assuming co-rotation. Since the precision of PTF lightcurves is far lower than \added{the precision of} the Kepler lightcurves, the only significant effect this assumption has in our analysis is on the estimate of the mass-ratio. If a star is rotating faster than the orbital period, the mass-ratio ($q$) is overestimated. We quantify this by simulating a typical EL CVn lightcurve with a primary star which is rotating 2 and 4 times faster than the orbital period, while keeping all other parameters the same. Fitting these lightcurves with the model assuming co-rotation, results in values for $q$ of 0.02 and 0.10 higher the initial value of $q=0.17$.
All other lightcurve parameters do not change significantly. We therefore conclude that for mildly faster-than-synchronous rotating primary stars ($P_\mathrm{rot}/P_\mathrm{orb}>0.5$), the effect on the mass-ratio is similar or smaller than the statistical uncertainty on the mass-ratio. If the primary star is \replaced{rotating faster}{rapidly rotating}, the mass-ratio is overestimated.

This overestimate propagates through to the rest of our parameter estimates; the average density of the primary is overestimated, and therefore the mass of the primary underestimated (Fig. \ref{fig:density}), and the semi-major axis is overestimated. However, the effect is small since these parameters only weakly depend on the mass-ratio (see Equations \ref{eq:rho} and \ref{eq:sma}). For a large part of our sample, we do not have any radial velocity amplitudes, and for these systems we rely on the mass-ratio to calculate the mass of the pre-He-WD. As mentioned in Section \ref{sec:calcMR}, the mass of the pre-He-WD depends on the mass-ratio to the third power. This, \replaced{combined with}{and a combination of} a high statistical and systematic uncertainty on the mass-ratio, makes the  calculations of $M_2$ (without a radial velocity amplitude) unreliable.

To check if the A/F-stars are rotating faster than synchronous, we compare the rotation periods to the orbital period for stars in known EL CVn systems. The orbital period of the main sequence star has been determined for five Kepler EL CVn binaries by measuring the projected rotational velocity ($v \sin i$): 1.79(60)\,d, 0.79(14)\,d, 5.0(2.4)\,d, and 1.71(62)\,d; \citep{2015ApJ...815...26F} and \citep[0.93\,d,][]{2013A&A...557A..79L}.
In addition, the rotational period has tentatively been identified from a frequency analysis for KOI-81 \citep[$0.48$\,d][]{2015ApJ...806..155M}, KOI-74 \citep[0.59\,d][]{2012MNRAS.422.2600B}, KOI-1224 \citep[3.49\,d][]{2012ApJ...748..115B}, and KIC-8262223 \citep[0.62\,d][]{2017ApJ...837..114G}. All rotational periods are of the same order as the orbital period. A detailed comparison of the rotational and orbital period shows that most stars rotate faster than synchronous, but in one case the rotation period is longer than the orbital period. For three cases (all from \citealt{2015ApJ...815...26F}) the rotation period is consistent with the orbital period of the binary, but since the uncertainties on the rotational periods are large, it is difficult to say if they are synchronised. The data therefore indicates that at least some (or maybe most) of the A/F-stars are not synchronised with the orbital period.

\begin{figure*}
\includegraphics{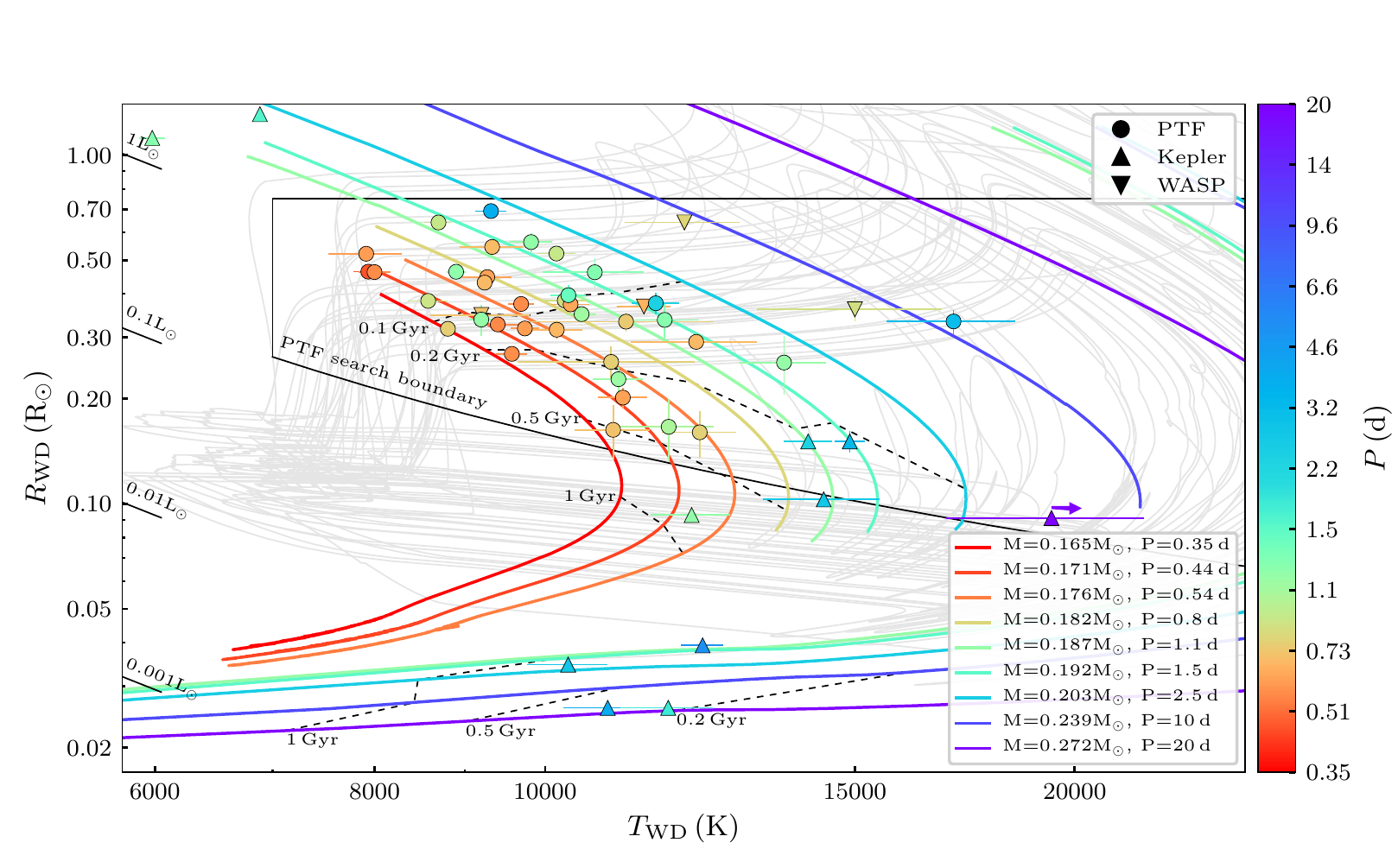}
\caption{Temperature versus radius of the pre-He-WDs, with the colours indicating different orbital periods. The coloured lines indicate evolution tracks by \citet{2013A&A...557A..19A} for different masses. The dots indicate pre-He-WDs from this study, triangles are other low mass pre-He-WDs in EL CVn systems; upward pointing triangles indicate Kepler discoveries, downward pointing triangles indicate SWASP discoveries (for references, see Sec. \ref{sec:intro}). During the evolution of pre-He-WDs, multiple hydrogen shell flashes can occur, indicated as grey lines. The tracks before the first H-flash and after the last H-flash are shown as coloured lines. Isochrones are shown as dashed lines, counted from the end of mass-transfer in the binary. The solid black line indicates the approximate detection limits, estimated by assuming a $T_1=$7000\,K, $R_1$=1.5\,\Rsun\ primary star. The bottom boundary is set by an eclipse depth of 0.03\,mag in $R$-band, the top boundary is set by the requirement of a flat-bottom eclipse ($R_1 > 2 R_2$), and the left limit is set by the requirement that the flat-bottom eclipse is deeper than the secondary eclipse ($T_1<T_2$).
}
\label{fig:TR}
\end{figure*}

There is however an important difference between the PTF sample and the sample of EL CVn systems with known rotation periods of the primary (all found by Kepler). The relative size of the A/F-star ($r_1$) is a factor of $\unsim3$ larger in the PTF sample, which strongly affects the synchronisation timescale of the star \citep[$\propto r_1^{-8.5}$,][]{1977A&A....57..383Z}. We used the equation by \citet{1977A&A....57..383Z} and tabulated values for $\mathrm{E_2}$ from \citet{2004A&A...424..919C} to calculate the synchronisation timescale for each of the EL CVn systems in our sample. This shows that the synchronisation timescale of the A/F-type star is less than 10\,Myr in 20 systems, and less than 100\,Myr for 32 systems. For these 32 systems, the synchronisation time is shorter than the time since mass transfer (0--260\,Myr, see Fig. \ref{fig:TR}). The remaining 4 systems (with the smallest values for $r_1$) have synchronisation timescales that are significantly longer than the estimated age. Based on this theoretical prediction, we can assume that that most of the A/F-stars are rotating synchronously. Whether this is actually the case requires an independent measurement of the rotation period.

\begin{figure*}
\includegraphics{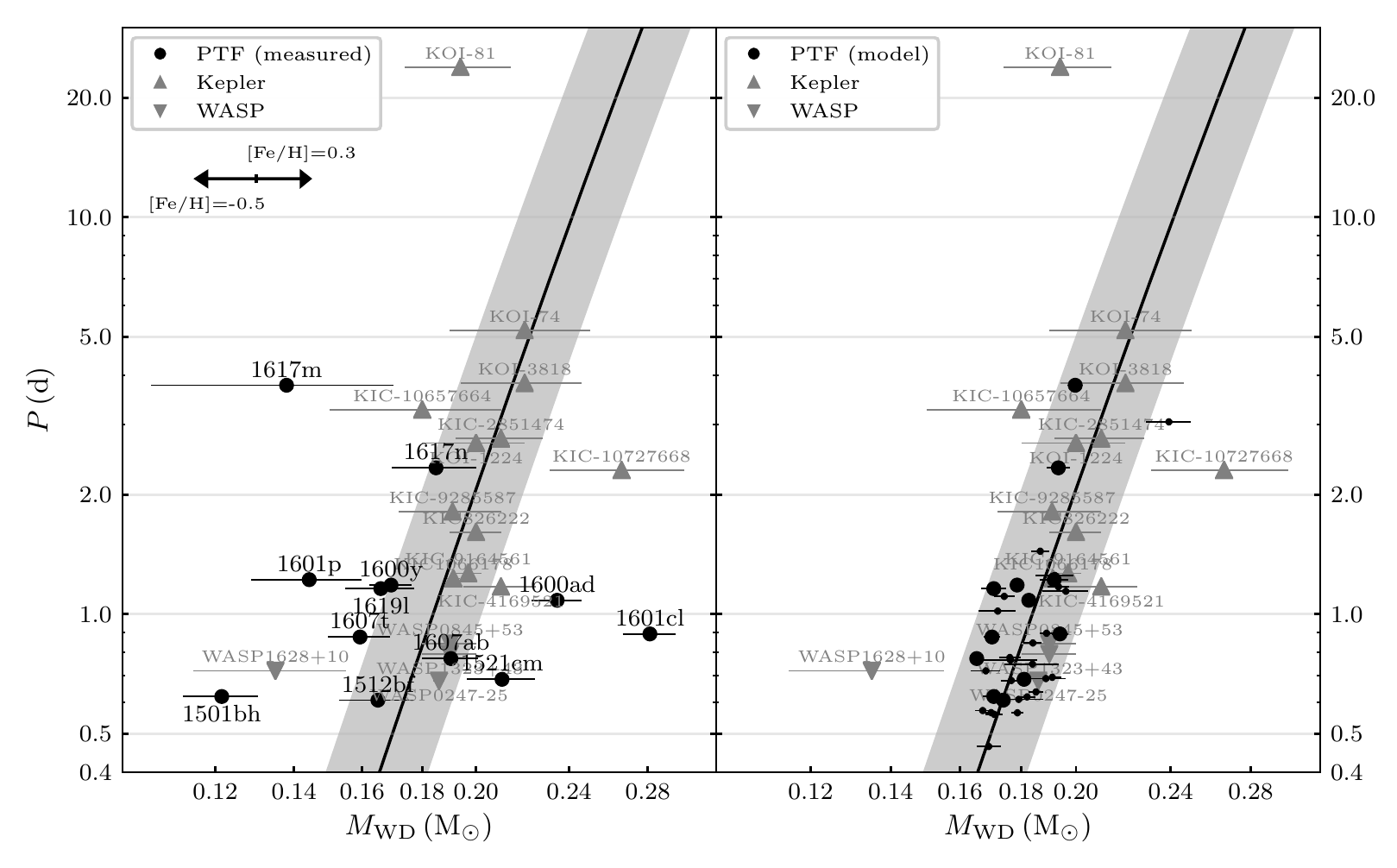}
\caption{Pre-He-WD mass versus the orbital period of the binary. The left panel shows the measured mass (using the radial velocity amplitude) of the PTF sample, the right panel shows the mass derived from the radius and temperature models (see Fig. \ref{fig:TR}). Large points indicate PTF systems with a radial velocity measurement, small points indicate systems without a radial velocity measurement \added{(not shown in the right panel because of the high uncertainties)}. Grey triangles indicate mass-period measurements of other EL CVn systems; upward pointing triangles indicate Kepler discoveries, downward pointing symbols indicate SWASP discoveries (for references, see Sec. \ref{sec:intro}). The black line shows the period-mass relation by \citet{2011ApJ...732...70L}, and the shaded area indicates a 10~per cent uncertainty on this relation. }
\label{fig:PM}
\end{figure*}

\subsection{Binary evolution and stellar parameters}\label{sec:evolution}

In the canonical formation channel of EL CVn binaries \citep[e.g.][]{2017MNRAS.467.1874C}, two main-sequence stars of similar mass are born at a short orbital period of a few days. The more massive star evolves faster and increases in radius. Before it ascends the red giant branch (RGB), it fills its Roche lobe and starts stable mass-transfer to the lower mass secondary star. This process continues until almost the complete outer envelope is transferred \citep[identified as ``R CMa''-type binaries, e.g.][]{2016AJ....151...25L}. The remnant of the initially more massive star has become a pre-white dwarf with a helium core and a thick hydrogen envelope ($\approx 0.01-0.04$\,\Msun, see \citealt{2016A&A...595A..35I,2017MNRAS.467.1874C}). The accretor has become a rejuvenated main-sequence star of spectral type A or F which dominates the luminosity of the system. If present in a specific stellar population as e.g. found in clusters, such a system would be identified as a `blue straggler'. If the orbital inclination is such that it shows eclipses, we identify it as an EL CVn binary.

The structure and evolution of pre-He-WDs have been extensively studied, as they also occur at more advanced binary evolutionary stages with either a white dwarf \citep{1995MNRAS.275..828M} or a neutron star \citep{2005ASPC..328..357V} as companions. Modelling of the formation process of binaries with a low mass pre-He-WD \citep{2013A&A...557A..19A,2014A&A...571L...3I,2014A&A...571A..45I,2016A&A...595A..35I,2017MNRAS.467.1874C} shows that there are strong correlations between the binary orbital period, and the mass, temperature, radius and age of the pre-He-WD. First, higher mass pre-He-WDs are formed at longer orbital periods. This is a direct result of the mass accretion process. This relation was already found in pre-He-WD -- neutron star binaries and has been parametrised by \citet{2011ApJ...732...70L}. Binary evolution studies by \citet{2014A&A...571A..45I} and \citet{2017MNRAS.467.1874C} also predict this P-M relation for EL CVn binaries. The latter shows that the relation between orbital period and mass is very robust, but at the low-mass end of the relation (0.16--0.20\Msun) there is some spread.

Pre-He white dwarfs (of a given mass) are also predicted to follow a particular evolutionary track, corresponding to a particular combination of radius and temperature as a function of age. The temperature and radius are directly related to the envelope mass and core mass of the white dwarf. Directly after the mass accretion process ends, the pre-He-WD is large ($\gtrsim 0.5$\,\Rsun) and has a low ($\lesssim 8000$\,K) surface temperature. While the hydrogen envelope is slowly being consumed by shell burning, the pre-He-WD shrinks and increases in temperature while maintaining an approximately constant luminosity (this phase is often referred to as the constant luminosity phase). When almost the entire envelope has been consumed, the pre-He-WD starts to cool down while the radius keeps decreasing (the cooling track). At the beginning of the cooling track, multiple hydrogen shell flashes \citep[H-flash, e.g.][]{1998A&A...339..123D}
can occur in the more massive pre-He-WDs. These flashes briefly increase the temperature and radius of the star, after which the white dwarf settles back on the cooling track. The exact mass boundary at which this starts to occur is uncertain. Models by \citet{2013A&A...557A..19A} show shell flashes for masses above $\unsim$0.18\Msun, while \citet{2014A&A...571A..45I} put this boundary at $\unsim$0.21\Msun.

\deleted{In }Fig.~\ref{fig:TR} shows the temperature versus radius of the pre-He-WD, with the colour of the points indicating the orbital period of the system. This shows that the temperatures and radii we find are consistent with predictions for pre-He-WDs in the constant luminosity phase, and before the occurrence of any H-flash. While some of the measurements are also consistent with pre-He-WDs undergoing an H-flash \added(grey lines), the short time spent in this phase makes this extremely unlikely. For the PTF sample, the orbital period (indicated by the colours) follows the same trend as the models, with long period systems containing larger and hotter pre-He-WDs. To test if the data match the models, we interpolate the models in orbital period, which allows us to test directly how well the radius and temperature match the model for a given orbital period. The fraction of measurements within 1, 3, and 5 standard deviations is 25, 75 and 86 per cent. Given the fact that we interpolate, and the uncertainties on radius and temperature could contain some systematic uncertainties, we conclude that most of the measurements agree with the models. This comparison to the models also allows us to infer the time since the end of mass transfer, which ranges from 0 to 260\,Myr with an average of 110\,Myr.


\deleted{In }Fig.~\ref{fig:PM} shows the orbital period of the binary versus the mass of the pre-He-WD. The left panel shows the measured values, while the right panel shows the expected masses using the models inferred from the measured temperature and radius (Fig.~\ref{fig:TR}). The measured values indicate that all pre-He-WD are low mass systems, but the PTF sample scatters around the model predictions. The right panel shows that if the radius and temperature measurements and models are used to derive the mass, the results fall within 10 per cent of the prediction of the mass-period relation for pre-He-WDs.

There are two possible explanations for this discrepancy. Either we have underestimated the uncertainties on the mass measurements, or there is some additional intrinsic spread on the predicted mass versus period, radius, and temperature not properly modelled. There are a number of possible systematic uncertainties that could affect the mass determination. First of all, we have assumed that the A/F-star is a regular main sequence star with a solar metallicity to estimate its mass (see Fig.~\ref{fig:density}). In Fig.~\ref{fig:PM} we have indicated how the mass estimate changes if we assume a different metallicity. If the real metallicity is lower than assumed, the mass is overestimated. This could explain a part of the inconsistency with the theory, but extreme metallicities would be needed to explain the largest outliers. Since thick disk systems generally have a lower metallicity, the masses for these systems could be overestimated. However, the thick disk systems do not show any particular trend, indicating that this assumption may not be the dominant cause of the inconsistency with the model predictions.

Another possibility is that we have underestimated our measurement uncertainties. The mass of the pre-He-WD is mostly determined by the radial velocity measurement. As shown in Fig.~\ref{fig:RVcurves} in the Appendix, we need to add an additional uncertainty to the formal uncertainties in order to explain all variance in the radial velocity measurements. For PTFS1601cl (one of the outliers), where we did not obtain radial velocity measurements at the quadratures, small systematic offsets between measurements can have a large impact on the radial velocity amplitude. We did check the radial velocity amplitude measurements of PTFS1512bf by obtaining a few high-resolution spectra with the Echellette Spectrograph and Imager (ESI) on the Keck telescope. The resulting radial velocity amplitude measured from these spectra is consistent with the result from the IDS spectra, which leads to the conclusion that uncertainties due to an unstable detector are most likely very small.

An alternative explanation is that there is some intrinsic variance between mass and period, radius and temperature. For example, \citet{2016A&A...595L..12I} shows that assumptions about rotation, diffusion and metallicity give different results when modelling the mass, radius and temperature of pre-He-WDs. The magnitude of the effect is estimated to be low, about 10 per cent. This would be enough to explain the variance in the right panel but cannot explain the outliers on the left panel.

To solve this ambiguity, a measurement of both the main-sequence and pre-He-WD radial velocity is needed. This allows the mass of both stars to be calculated by only using Kepler's law (combined with the period and inclination measurement from the lightcurve). This has been done for SWASP J0247-25 \citep{2013Natur.498..463M}, KOI-81 \citep{2015ApJ...806..155M}, KIC-10661783, \citep{2013A&A...557A..79L}, and KIC-8262223 \citep{2017ApJ...837..114G}. For SWASP J0247-25, KIC-10661783, and KIC-10661783 the mass of the pre-He-WD agrees well with the P-M relation, but for KOI-81, the mass is significantly lower (0.10\Msun) than the P-M relation predicts. This hints that there is more scatter in the P-M relation than models estimated.

\subsection{Galactic population and space density}

Using stellar evolution and population synthesis codes, \citet{2017MNRAS.467.1874C} predict a space density of $4$--$10\,\times\,10^{-6}\,\mathrm{pc^{-3}}$ for EL CVn binaries (including non-eclipsing ones) with orbital periods less than 2.2\,d. In addition, they predicted that EL CVn binaries should mainly be found in young stellar populations, and therefore be more abundant in the thin disk. We use the Galaxy model based on SDSS data by \citet{2008ApJ...673..864J} to estimate how many EL CVn binaries we would expect to see given this space density and in what ratios between thin disk, thick disk and halo. We populate our model Galaxy with stars with absolute magnitudes according to a normal distribution with a mean and standard deviation of $R=2.46\pm0.54$\,mag, values determined from our sample of 36 systems. We simulate the PTF coverage by using (overly) simple requirements: $|b|>15$, $\delta>0$, $13.5<R<16$ (see Fig.\ \ref{fig:overview}). We ignore the Galactic Plane because these fields tend to be observed only $\unsim50$ times. The minimum number of epochs in our uncovered sample is 58, indicating that at least 58 observations are needed to identify an EL CVn \added{binary}. Using the 58 epoch limit, we derive an effective coverage of 32.8 per cent for the remaining area. We also correct for the requirement that the systems must be eclipsing. This decreases the number of observable EL CVn systems by a factor of $0.307$; determined from our sample using radii and inclination. Even if the binaries are eclipsing, if the pre-He-WD is too small (and thus old), we would not find it in the PTF data. To correct for this, we assume a lifetime of EL CVn binaries of 1\,Gyr (the main sequence lifetime of a 2\,\Msun\ star) and compare this to the typical age of PTF EL CVn binaries (0--260\,Myr). We therefore assume that PTF can only detect 26 percent of all EL CVn \replaced{binaries}{stars}.

According to the model and the assumed selection criteria, 26 per cent of the PTF sample should be from the thin disk. If \added{we} assume our model is correct, there is only a 1.8 per cent \added{chance} ($\sum_{n=7}^{12} \binom{12}{n}\, 0.26^n\, [1-0.26]^{12-n}$) to find $\geq 7$ thin disk systems out of a total of 12 EL CVn systems. If any of the ambiguous cases are from the thin disk, this probability drops well below 1 per cent. This indicates that our model is unlikely to be correct, and confirms that EL CVn systems are more abundant in the thin disk compared to the average stellar population, as was already suggested by \citet{2017MNRAS.467.1874C}.

Using the model and estimated PTF detection efficiency, we also predict that we should have found $\unsim300$--$750$ EL CVn systems, a factor of $\unsim10-25$ higher than we actually recovered. This could simply be because we are over-predicting the contributions of the thick disk and halo. However, even if we assume a factor of 4 higher contribution from the thin disk (to bring the model in line with with the ratio of thin to thick disk systems), the model still predicts at least a factor of $5-12$ more EL CVn systems than we found. Another uncertain estimate which could explain the discrepancy is the assumed efficiency of PTF in finding EL CVn binaries. The PTF observing cadence and coverage is highly inhomogeneous, and the assumptions we have used are very simple approximations. Assuming that we can find all EL CVn systems observed more than 58 times and are younger than 200\,Myr is overly optimistic, and could explain the discrepancy of a factor of 5 (or more).

The inhomogeneity of the PTF dataset makes it difficult to do a proper study of the Galactic distribution and space density of EL CVn systems. We do find tentative results that EL CVn systems occur more often in the thin disk, as was predicted by \citealt{2017MNRAS.467.1874C}. We also find that the space density is at the lower bound, or even below the prediction of 4--10 $\times 10^{-6}\,\mathrm{pc^{-3}}$. To properly measure the properties of the population of EL CVn systems, a larger sample of EL CVn binaries is needed, preferably from a more uniform variability survey.

\subsection{Comparison with the SWASP sample}
\begin{figure}
\includegraphics{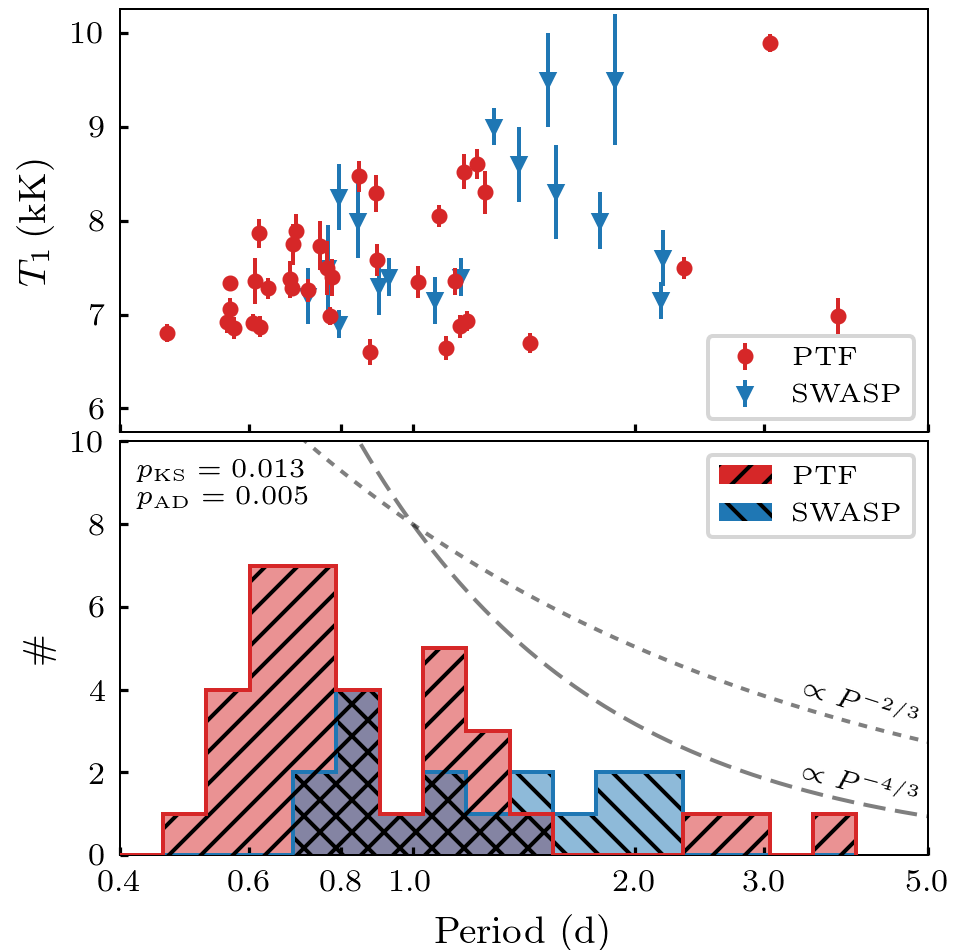}
\caption{(\textit{top}) The temperature of the A/F-star versus the orbital period of the EL CVn system. (\textit{bottom}) The distribution of orbital periods of the PTF sample (red) and the SWASP sample (blue). In the top left, the result of the KS-test and AD-test are shown (see text). 
The top panel shows that there is a strong correlation between orbital period and temperature, which is a result of the binary evolution process (see text). The histograms show that the PTF data is more biased to the short period systems compared to the SWASP sample. It also shows a possible gap at periods of 1 day, caused by a detection bias against these systems. For comparison, we also plotted the detection probability of an eclipsing population with well-sampled lightcurves ($\propto P^{-2/3}$, dotted line), and for lightcurves with a limited amount of epochs ($\propto P^{-4/3}$, dashed line). }

\label{fig:periods}
\end{figure}

To better understand the biases of our survey, we compare the PTF sample to the sample found by SWASP \citep{2014MNRAS.437.1681M}. While we both have used the lightcurve characteristics to identify EL CVn binaries, there are some intrinsic differences between the surveys, and therefore different biases in finding EL CVn systems. 
The most obvious difference between the surveys is the magnitude range; SWASP probes magnitudes between 9--13\,mag, while our sample is fainter, between 13.5 and 16\,mag. A second major difference is the cadence and the number of epochs in a lightcurve; PTF lightcurves have an irregular cadence and a low number of epochs ($\unsim$100) compared to the regular cadence and better-sampled SWASP lightcurves ($\unsim4000$--$13000$ epochs).

There are indeed differences between the two samples. First , the distance range for the SWASP sample is 100--1200 pc, while the PTF sample reaches 1200--5000\,pc. This is expected given the different magnitude range of the two surveys. We therefore also expect to find relatively more thick disk and halo systems compared to thin disk systems \added{in the PTF sample}. However, using the one-sided Fisher's exact test \citep{Fisher}, we find no significant difference between the relative number of thin disk systems. This is consistent with our finding that EL CVn systems are more numerous in the thin disk compared to the average stellar population and explains why at larger distances it is still the most dominant population.

The two samples are also different with regards to orbital period and temperature of the A/F-star (which dominates the luminosity), see Fig \ref{fig:periods}. We performed a Kolmogorov--Smirnov (KS) test and Anderson--Darling (AD) test \citep[e.g. Sec. 3.1 in][]{2012msma.book.....F} to compare the distribution of orbital periods. Both tests show that it is unlikely that the samples are drawn from the same distribution ($p_\mathrm{KS}=0.013$ and $p_\mathrm{AD}=0.005$). The histogram in Fig. \ref{fig:periods} shows that we find more short orbital period systems and the top panel in Fig. \ref{fig:periods} shows that at short orbital periods, the temperature of the primary star is low. This correlation can be understood because the mass of A/F-star is correlated with the orbital period. High mass main-sequence stars (2\,\Msun) begin their main-sequence lifetime at a temperature of 9500\,K, and cool down to 7500\,K towards the end of their main sequence lifetime (see Fig. \ref{fig:density}). Main sequence stars of 1.3\,\Msun\ start at a temperature of 7500\,K and only cools by 500\,K during their time on the main sequence. Therefore, PTFs sensitivity for lower luminosity (lower temperature) EL CVn systems (partially) explains why the PTF sample contains more short period systems. 

 A second explanation is that, because of the sparse sampling, it is harder to find long period systems with PTF compared to SWASP. Both surveys use eclipses to find EL CVn binaries, and therefore are biased towards short period systems ($\mathrm{Prob_{ecl}} \propto R_1 P^{-2/3}$). In addition, short period systems spend a larger fraction of their orbit in eclipse ($\mathrm{\tau_{ecl}} \propto R_1 P^{-2/3}$). This does not bias the SWASP search since the lightcurves are well sampled. \replaced{With}{In} PTF however, a lack of observations during the eclipse can hinder the identification of a system as an EL CVn binary.

The difference between the two samples shows that selection effects make it difficult to determine the intrinsic properties of EL CVn binaries. To do so requires an integrated approach: stellar evolution and population synthesis models should be used to simulate a sample of EL CVn binaries, which are then `observed' by simulating the variability survey which was used to find the real sample. Such a calculation is difficult given the in-homogeneity of the PTF sample, and beyond the scope of this work.

\subsection{Pulsations}
Pre-He-WDs are predicted to exhibit both p- and g-mode pulsations \citep[e.g.][]{2014A&A...569A.106C,2016A&A...585A...1C,2016A&A...588A..74C,2016A&A...595L..12I}. Pulsations have been found in two of the SWASP EL CVn binaries: WASP 0247-25 \citep{2013Natur.498..463M,2017ApJ...847..130I} and WASP 1628+10 \citep{2014MNRAS.444..208M}. The pulsation periods are 5--10 minutes and the amplitudes $\unsim$1--2 per cent of the pre-He-WD luminosity. Models of pre-He-WDs predict that in a large area in T--$\log g$ space, pre-He-WDs should feature pulsations \citep[Fig. 10 in ][]{2016A&A...588A..74C,2016A&A...595L..12I}. Many of the pre-He-WDs \replaced{in}{from} the PTF sample lie in this region, making them useful to test the general predictions for pulsation theory. In addition, because stellar parameters can be measured very precisely, a pulsating pre-He-WD\deleted{s} in an eclipsing binary is extremely useful to test evolutionary and seismic models in great detail \citep[e.g.][]{2017ApJ...847..130I}.

Unfortunately, the very sparse sampling of the PTF lightcurves makes it very difficult to identify such pulsations. We did attempt to find pulsations by using a Lomb-Scargle algorithm (\citealt{1976Ap&SS..39..447L,1982ApJ...263..835S}, implementation by \citealt{2015ApJ...812...18V}) on the residuals of the lightcurves. We found periodic behaviour in the residuals at predicted periods of $\unsim$10 minutes for a number of the systems, but because of the sparse sampling and low amplitude, it is difficult to determine if these are real or not. High cadence follow-up photometry is needed to establish the reality of these pulsations.

\section{Summary and conclusion} \label{sec:summary}
In this paper, we report the discovery and analysis of 36 new EL CVn systems extracted from \deleted{the} Palomar Transient Factory data. With this sample, we more than double the number of known EL CVn systems. To find the EL CVn systems we used machine learning classifiers to make a pre-selection of candidates. This has proven to be an efficient method to minimise the number of lightcurves that have to be visually inspected.

The radii (0.16--0.7\,\Rsun) and temperatures (8000--17000\,K) of the pre-He-WDs in the sample indicate they are all young systems in the ``constant luminosity'' phase (0--250\,Myr) of their evolution. The masses of the pre-He-WDs are all low (<0.3\Msun), but our measurements show a large spread around the predicted mass-period relation, which remains unexplained. If we use the measured radii and temperatures combined with models, we do find masses consistent with the mass-period relation. This discrepancy is either due to systematic or underestimated uncertainties in our measurements, or there is more variance in the masses than the stellar evolution models predict. This problem can be resolved by obtaining more accurate radial velocity measurements (ideally for both stars in the binary to obtain an independent mass ratio measurement), and by more extensively testing the effect on the mass-period relation of e.g. different metallicities and rotation rates.

Although a detailed population study is difficult with the PTF dataset, we find that EL CVn binaries occur more often in the thin disk than an average Galactic stellar population. In addition, we find that the space density is most likely lower than the predicted value of $4$--$10\,\times\,10^{-6}\,\mathrm{pc^{-3}}$. To properly determine the properties of the EL CVn population, a more systematic search combined with stellar and Galactic modelling is required.

This new sample of young pre-He-WD\added{s} will be useful to put many theoretical models to the test, including stellar structure models for low mass white dwarfs, pulsation models, and binary evolution models. In addition, the methods we have used to identify EL CVn systems can easily be adapted to find other rare types of variable stars and these (and similar machine learning methods) will be vital to fully utilise (future) variability surveys like
ZTF \citep{2014htu..conf...27B}, 
NGTS \citep{2017arXiv171011100W},
GOTO \citep{GOTO}, 
BlackGEM \citep{2015ASPC..496..254B}, 
\textit{TESS} \citep{2015JATIS...1a4003R},
\textit{PLATO} \citep{Rauer2014},
and LSST \citep{lsst}.

\section*{Acknowledgements}
We thank the referee, Simon Jeffery, for thoroughly reading the manuscript and providing us with useful comments and suggestions.

We thank Adam Miller, Fabian Gieseke and Tom Heskes for many useful discussions about machine learning classification. We thank Tom Marsh for the use of \textsc{lcurve}. We thank Luc Hendriks for suggesting to use \textsc{XGBoost}.

JvR acknowledges support from the Netherlands Research School of Astronomy (NOVA) and Foundation for Fundamental Research on Matter (FOM), and also the California Institute of Technology were a large part of this work was conducted. JvR and PJG thank the University of Cape Town for their hospitality: this work was finalised while visiting UCT, supported by the NWO-NRF Bilateral Agreement in Astronomy. This research was supported in part by the National Science Foundation under Grant No. NSF PHY-1125915 and PJG thanks the Kavli Institute for Theoretical Physics at the University of California, Santa Barbara for a productive stay. 

The Intermediate Palomar Transient Factory project is a scientific collaboration among the California Institute of Technology, Los Alamos National Laboratory, the University of Wisconsin, Milwaukee, the Oskar Klein Center, the Weizmann Institute of Science, the TANGO Program of the University System of Taiwan, and the Kavli Institute for the Physics and Mathematics of the Universe. 

Funding for the Sloan Digital Sky Survey IV has been provided by the Alfred P. Sloan Foundation, the U.S. Department of Energy Office of Science, and the Participating Institutions. SDSS-IV acknowledges support and resources from the Center for High-Performance Computing at the University of Utah. The SDSS web site is www.sdss.org.


The Pan-STARRS1 Surveys (PS1) and the PS1 public science archive have been made possible through contributions by the Institute for Astronomy, the University of Hawaii, the Pan-STARRS Project Office, the Max-Planck Society and its participating institutes, the Max Planck Institute for Astronomy, Heidelberg and the Max Planck Institute for Extraterrestrial Physics, Garching, The Johns Hopkins University, Durham University, the University of Edinburgh, the Queen's University Belfast, the Harvard-Smithsonian Center for Astrophysics, the Las Cumbres Observatory Global Telescope Network Incorporated, the National Central University of Taiwan, the Space Telescope Science Institute, the National Aeronautics and Space Administration under Grant No. NNX08AR22G issued through the Planetary Science Division of the NASA Science Mission Directorate, the National Science Foundation Grant No. AST-1238877, the University of Maryland, Eotvos Lorand University (ELTE), the Los Alamos National Laboratory, and the Gordon and Betty Moore Foundation.

This publication made use of data products from the Two Micron All Sky Survey, which is a joint project of the University of Massachusetts and the Infrared Processing and Analysis Center/California Institute of Technology, funded by the National Aeronautics and Space Administration and the National Science Foundation.

Based on observations made with the NASA Galaxy Evolution Explorer. GALEX is operated for NASA by the California Institute of Technology under NASA contract NAS5-98034. 

This publication made use of data products from the Wide-field Infrared Survey Explorer, which is a joint project of the University of California, Los Angeles, and the Jet Propulsion Laboratory/California Institute of Technology, funded by the National Aeronautics and Space Administration.

This research has made use of the VizieR catalogue access tool, CDS, Strasbourg, France. The original description of the VizieR service was published in A\&AS 143, 23.

This research has made use of the SVO Filter Profile Service (http://svo2.cab.inta-csic.es/theory/fps/) supported from the Spanish MINECO through grant AyA2014-55216.

This research made use of Scikit-learn \citep{scikit-learn}.

This research made use of Astropy, a community-developed core Python package for Astronomy \citep{2013A&A...558A..33A}.

IRAF is distributed by the National Optical Astronomy Observatory, which is operated by the Association of Universities for Research in Astronomy (AURA) under cooperative agreement with the National Science Foundation \citep{1993ASPC...52..173T}.




\bibliographystyle{mnras}
\bibliography{bibfiles/ELCVn_misc,bibfiles/ELCVn_other,bibfiles/ELCVn_ML,bibfiles/ELCVn_theory} 



\appendix

\section{Additional tables and figures}
\clearpage

\begin{table}
\centering
\caption{List of a features used by the machine learning classifiers.}
\label{tab:featurelist}
\begin{tabular}{p{2.5cm} p{5.5cm} }
\multicolumn{2}{l}{PTF variability} \\
\hline
wRMS (mag) & weighted root-mean-square of the PTF lightcurve \\
skew (mag) & skewness of the PTF lightcurve \\
medAbsDev (mag) & median absolute deviation of the PTF lightcurve \\
StetsJ & Stetson-J statistic of the PTF lightcurve \\
StetsK & Stetson-K statistic of the PTF lightcurve \\
Neumann & the Von Neumann ratio statistic of the PTF lightcurve \\
MBR & median buffer range: the fraction of points more than 20\% of the lightcurve amplitude from the weighted mean magnitude, divided by total number of epochs. \\
AMBS\{1,2,3\} & The fraction of lightcurve points that are \# standard deviation above the mean magnitude. \\
BMBS\{1,2,3\} & The fraction of lightcurve points that are \# standard deviation below the mean magnitude. \\
prange\{\#\} (mag) & range containing \{90,80,65,50,35,20\} per cent of the data. \\
percentile\{\#\} & \#th percentile minus the median of the PTF lightcurve, divided by prange90 , with \# in \{5,10,17.5,25,32.5,40,60,67.5,75,82.5,90,95\}. \\
\vspace{0.1cm} \\
\multicolumn{2}{l}{PAN-STARRS colours} \\
\hline
PSr (mag) & Pan-STARRS $r$ - median of the lightcurve\\
PSgr (mag) & Pan-STARRS $g-r$\\
PSri (mag) & Pan-STARRS $r-i$\\
PSiz (mag) & Pan-STARRS $i-z$\\
PSzy (mag) & Pan-STARRS $z-y$\\
\vspace{0.1cm} \\
\multicolumn{2}{l}{2MASS colours} \\
\hline
J (mag) & 2MASS $J$ - median of the lightcurve\\
JH (mag) & 2MASS $J-H$\\
HK (mag) & 2MASS $H-K$\\
\hline
\end{tabular}
\end{table}

\begin{table}
\centering
\caption{Overview of the nights at the INT with the IDS.}
\label{tab:specobs}
\begin{tabular}{lllll}
Date & Grating & CCD & seeing (\arcsec) & weather \\
\hline
2016-09-07 & R632V & RED+2 & 0.6 & excellent \\
2016-09-08 & R632V & RED+2 & 0.7 & excellent \\
2016-09-09 & R632V & RED+2 & 0.6 & excellent \\
2016-09-10 & R632V & RED+2 & 0.7 & good \\
2016-09-11 & R632V & RED+2 & 0.6--1.0 & good \\
2016-09-12 & R632V & RED+2 & 0.7--1.0 & good \\
2016-09-13 & R632V & RED+2 & 0.8--1.2 & good \\
2016-09-14 & R632V & RED+2 & 1.0 & good \\
\hline
2016-12-14 & R900V & RED+2 & 0.8--1.4 & good \\
2016-12-15 & R900V & RED+2 & 1.4 & ok \\
\hline
2017-01-09 & R900V & RED+2 & 0.8 & good \\
2017-01-10 & R900V & RED+2 & 1.2--2.6 & ok--bad \\
\hline
2017-03-10 & R900V & EEV10 & 1.5--3.0 & bad \\
2017-03-11 & R900V & EEV10 & -- & clouds \\
2017-03-12 & R900V & EEV10 & 2--4 & bad \\
2017-03-13 & R900V & EEV10 & 1.5 & bad \\
\hline
\end{tabular}
\end{table}

\begin{table}
\centering
\caption{The temperatures of the A/F-star \added($T_1$) and pre-Helium white dwarf \added($T_2$) determined from the spectral energy distribution of the binary stars, see Fig.~\ref{fig:SEDcurves}. The $E(B-V)$ values are taken from \citet{1998ApJ...500..525S,2011ApJ...737..103S}, with an uncertainty of 0.034 \citep[as in][]{2011MNRAS.418.1156M}. The `RMS' column indicates the additional uncertainty added to account for all variance, which is achieved by the parameter $f$ in Equation~\ref{eq:SEDfit}.}
\label{tab:SEDfit}
\begin{tabular}{lllll}
ID & $T_1$ (K) & $T_2$ (K) & $E(B-V)$ & RMS \\
\hline
1600y & $6930 \pm 100$ & $8900 \pm 110$ & 0.047 & 0.05 \\
1600ad & $8050 \pm 120$ & $10490 \pm 200$ & 0.024 & 0.04 \\
1700do & $9890 \pm 110$ & $17100 \pm 1400$ & 0.015 & 0.03 \\
1600aa & $7890 \pm 190$ & $9300 \pm 400$ & 0.107 & 0.02 \\
1601p & $8600 \pm 160$ & $11700 \pm 500$ & 0.030 & 0.03 \\
1501bh & $6870 \pm 110$ & $11100 \pm 400$ & 0.035 & 0.03 \\
1601q & $8300 \pm 230$ & $10700 \pm 700$ & 0.081 & 0.03 \\
1601cl & $8280 \pm 200$ & $10100 \pm 300$ & 0.030 & 0.04 \\
1402de & $7870 \pm 150$ & $9300 \pm 300$ & 0.100 & 0.03 \\
1607aa & $8470 \pm 160$ & $10300 \pm 300$ & 0.087 & 0.04 \\
1607v & $7260 \pm 130$ & $10900 \pm 500$ & 0.090 & 0.03 \\
1607t & $6600 \pm 140$ & $8600 \pm 200$ & 0.009 & 0.10 \\
1607ab & $6980 \pm 100$ & $8810 \pm 80$ & 0.005 & 0.09 \\
1608ab & $7360 \pm 240$ & $7900 \pm 400$ & 0.037 & 0.05 \\
1612al & $7280 \pm 110$ & $10300 \pm 300$ & 0.039 & 0.01 \\
1512bf & $6920 \pm 90$ & $9740 \pm 180$ & 0.022 & 0.03 \\
1613s & $7350 \pm 140$ & $13700 \pm 800$ & 0.051 & 0.04 \\
1613u & $7340 \pm 70$ & $9690 \pm 160$ & 0.006 & 0.04 \\
1615ag & $7380 \pm 200$ & $10200 \pm 400$ & 0.093 & 0.05 \\
1615v & $6920 \pm 120$ & $9400 \pm 300$ & 0.030 & 0.03 \\
1515ay & $6800 \pm 100$ & $7930 \pm 150$ & 0.029 & 0.04 \\
1615w & $6690 \pm 110$ & $10300 \pm 200$ & 0.046 & 0.04 \\
1615ao & $7580 \pm 170$ & $8700 \pm 160$ & 0.070 & 0.03 \\
1615u & $7400 \pm 200$ & $12200 \pm 600$ & 0.069 & 0.06 \\
1616cr & $7060 \pm 120$ & $8000 \pm 170$ & 0.095 & 0.10 \\
1617n & $7500 \pm 110$ & $11600 \pm 400$ & 0.022 & 0.03 \\
1617m & $6990 \pm 190$ & $9320 \pm 190$ & 0.071 & 0.04 \\
1619l & $6870 \pm 120$ & $9200 \pm 150$ & 0.031 & 0.04 \\
1521ct & $8520 \pm 180$ & $9800 \pm 300$ & 0.090 & 0.02 \\
1621ax & $7340 \pm 170$ & $11800 \pm 700$ & 0.108 & 0.04 \\
1521cm & $7290 \pm 80$ & $9240 \pm 90$ & 0.051 & 0.02 \\
1622by & $7730 \pm 260$ & $11100 \pm 1400$ & 0.079 & 0.07 \\
1522cc & $6860 \pm 120$ & $9570 \pm 190$ & 0.042 & 0.06 \\
1622aa & $7490 \pm 290$ & $10900 \pm 1300$ & 0.160 & 0.11 \\
1622bt & $7750 \pm 220$ & $12200 \pm 1000$ & 0.061 & 0.07 \\
1723aj & $6640 \pm 130$ & $11000 \pm 400$ & 0.061 & 0.04 \\
\hline
\end{tabular}
\end{table}

\begin{table*}
\caption{The parameters of the lightcurve models shown in Fig.~\ref{fig:LCs}. This table shows the median and root-mean-square of final 5120 points in our MCMC chains. Note that these distributions are not normally distributed and parameters can be strongly correlated.}
\label{tab:lcstats}
\begin{tabular}{@{}l *{13}{r} }
ID & $P\,\mathrm{(d)}$ & $t_0\,\mathrm{(BMJD_{tdb})}$ & $i\,(\degree)$ & $q$ & $\langle r_1 \rangle$ & $\langle r_2 \rangle$ & $T_2/T_1$ & $\mathrm{absorb}_R$ & $\mathrm{absorb}_{g^\prime}$ & $\log(f_{R})$ & $\log(f_{g^\prime})$ & fill & $\rho\,(\rho_\mathrm{\sun})$ \\
Band & & & & & & & & & & & & & \\
\hline \noalign{\smallskip}
\noalign{\smallskip}
1600y & $ 1.1838920$  & $55570.2084$  & $      84.5$  & $      0.12$  & $     0.421$  & $     0.081$  & $      0.73$  & $       0.8$  & $       0.8$  & $      -2.1$  & $      -2.5$  & $     0.748$  & $     0.114$  \\
R+g & $ 0.0000008$  & $    0.0006$  & $       2.7$  & $      0.02$  & $     0.010$  & $     0.003$  & $      0.03$  & $       0.2$  & $       0.2$  & $       0.0$  & $       0.1$  & $     0.019$  & $     0.007$  \\
\noalign{\smallskip}
1600ad & $ 1.0840448$  & $56247.4677$  & $      86.5$  & $      0.11$  & $     0.328$  & $     0.063$  & $      0.80$  & $       1.3$  & $       1.0$  & $      -2.3$  & $      -2.4$  & $     0.575$  & $     0.292$  \\
R+g & $ 0.0000010$  & $    0.0006$  & $       2.2$  & $      0.03$  & $     0.008$  & $     0.003$  & $      0.03$  & $       0.5$  & $       0.5$  & $       0.0$  & $       0.1$  & $     0.025$  & $     0.019$  \\
\noalign{\smallskip}
1700do & $ 3.0507595$  & $55556.8044$  & $      87.4$  & $      0.33$  & $     0.178$  & $     0.025$  & $      0.77$  & $       2.8$  & $        $  & $      -2.3$  & $        $  & $     0.380$  & $     0.187$  \\
R & $ 0.0000281$  & $    0.0014$  & $       1.8$  & $      0.10$  & $     0.010$  & $     0.002$  & $      0.04$  & $       1.2$  & $        $  & $       0.0$  & $        $  & $     0.020$  & $     0.024$  \\
\noalign{\smallskip}
1600aa & $ 0.6934558$  & $56892.6368$  & $      78.7$  & $      0.29$  & $     0.391$  & $     0.128$  & $      0.94$  & $       3.6$  & $       3.2$  & $      -2.5$  & $      -2.6$  & $     0.801$  & $     0.362$  \\
R+g & $ 0.0000006$  & $    0.0003$  & $       0.8$  & $      0.05$  & $     0.007$  & $     0.004$  & $      0.01$  & $       0.8$  & $       0.8$  & $       0.2$  & $       0.4$  & $     0.026$  & $     0.021$  \\
\noalign{\smallskip}
1601p & $ 1.2215885$  & $57152.5232$  & $      84.0$  & $      0.21$  & $     0.274$  & $     0.056$  & $      0.84$  & $       1.4$  & $       1.1$  & $      -2.2$  & $      -2.3$  & $     0.534$  & $     0.358$  \\
R+g & $ 0.0000051$  & $    0.0010$  & $       3.1$  & $      0.09$  & $     0.019$  & $     0.005$  & $      0.03$  & $       1.0$  & $       1.1$  & $       0.0$  & $       0.3$  & $     0.036$  & $     0.059$  \\
\noalign{\smallskip}
1501bh & $ 0.6204144$  & $55097.3927$  & $      78.4$  & $      0.14$  & $     0.356$  & $     0.058$  & $      0.64$  & $       0.5$  & $       0.4$  & $      -2.1$  & $      -2.1$  & $     0.646$  & $     0.674$  \\
R+g & $ 0.0000005$  & $    0.0008$  & $       1.9$  & $      0.04$  & $     0.017$  & $     0.004$  & $      0.04$  & $       0.3$  & $       0.2$  & $       0.1$  & $       0.0$  & $     0.023$  & $     0.085$  \\
\noalign{\smallskip}
1601q & $ 1.2515054$  & $57190.1373$  & $      80.6$  & $      0.12$  & $     0.308$  & $     0.074$  & $      0.91$  & $       1.6$  & $       2.2$  & $      -2.3$  & $      -2.5$  & $     0.535$  & $     0.266$  \\
R+g & $ 0.0000051$  & $    0.0012$  & $       2.7$  & $      0.07$  & $     0.021$  & $     0.007$  & $      0.03$  & $       1.1$  & $       1.2$  & $       0.1$  & $       0.3$  & $     0.049$  & $     0.051$  \\
\noalign{\smallskip}
1601cl & $ 0.8917354$  & $56063.3087$  & $      82.9$  & $      0.11$  & $     0.475$  & $     0.101$  & $      0.95$  & $       2.3$  & $       1.9$  & $      -2.1$  & $      -2.3$  & $     0.830$  & $     0.142$  \\
R+g & $ 0.0000005$  & $    0.0004$  & $       2.9$  & $      0.02$  & $     0.011$  & $     0.004$  & $      0.02$  & $       0.8$  & $       0.7$  & $       0.0$  & $       0.0$  & $     0.021$  & $     0.009$  \\
\noalign{\smallskip}
1402de & $ 0.6189694$  & $55768.8152$  & $      87.0$  & $      0.17$  & $     0.405$  & $     0.119$  & $      0.91$  & $       2.5$  & $        $  & $      -2.9$  & $        $  & $     0.757$  & $     0.454$  \\
R & $ 0.0000011$  & $    0.0008$  & $       2.4$  & $      0.07$  & $     0.015$  & $     0.006$  & $      0.03$  & $       1.1$  & $        $  & $       0.3$  & $        $  & $     0.052$  & $     0.049$  \\
\noalign{\smallskip}
1607aa & $ 0.8463124$  & $56246.6579$  & $      84.4$  & $      0.14$  & $     0.375$  & $     0.079$  & $      0.92$  & $       2.2$  & $       3.7$  & $      -2.6$  & $      -2.6$  & $     0.678$  & $     0.314$  \\
R+g & $ 0.0000017$  & $    0.0007$  & $       3.6$  & $      0.03$  & $     0.017$  & $     0.005$  & $      0.03$  & $       1.0$  & $       0.9$  & $       0.3$  & $       0.2$  & $     0.024$  & $     0.035$  \\
\noalign{\smallskip}
1607v & $ 0.7198355$  & $55769.1206$  & $      82.7$  & $      0.13$  & $     0.447$  & $     0.039$  & $      0.66$  & $       2.0$  & $       2.2$  & $      -2.4$  & $      -2.3$  & $     0.808$  & $     0.256$  \\
R+g & $ 0.0000020$  & $    0.0014$  & $       5.7$  & $      0.04$  & $     0.037$  & $     0.007$  & $      0.08$  & $       1.0$  & $       1.0$  & $       0.1$  & $       0.0$  & $     0.038$  & $     0.048$  \\
\noalign{\smallskip}
1607t & $ 0.8759507$  & $56158.7102$  & $      76.6$  & $      0.09$  & $     0.417$  & $     0.085$  & $      0.74$  & $       0.7$  & $       0.5$  & $      -2.3$  & $      -2.3$  & $     0.712$  & $     0.220$  \\
R+g & $ 0.0000004$  & $    0.0004$  & $       1.0$  & $      0.02$  & $     0.009$  & $     0.003$  & $      0.03$  & $       0.2$  & $       0.2$  & $       0.0$  & $       0.0$  & $     0.020$  & $     0.014$  \\
\noalign{\smallskip}
1607ab & $ 0.7730986$  & $55151.7862$  & $      83.8$  & $      0.18$  & $     0.351$  & $     0.077$  & $      0.77$  & $       1.9$  & $       1.5$  & $      -2.3$  & $      -2.2$  & $     0.665$  & $     0.439$  \\
R+g & $ 0.0000002$  & $    0.0004$  & $       2.3$  & $      0.03$  & $     0.010$  & $     0.003$  & $      0.03$  & $       0.4$  & $       0.4$  & $       0.1$  & $       0.0$  & $     0.020$  & $     0.035$  \\
\noalign{\smallskip}
1608ab & $ 0.6101718$  & $57034.9178$  & $      86.8$  & $      0.07$  & $     0.390$  & $     0.146$  & $      0.90$  & $       0.4$  & $        $  & $      -3.0$  & $        $  & $     0.656$  & $     0.551$  \\
R & $ 0.0000014$  & $    0.0003$  & $       1.9$  & $      0.07$  & $     0.006$  & $     0.004$  & $      0.02$  & $       0.3$  & $        $  & $       0.3$  & $        $  & $     0.058$  & $     0.044$  \\
\noalign{\smallskip}
1612al & $ 0.6369260$  & $55782.6928$  & $      86.8$  & $      0.12$  & $     0.322$  & $     0.103$  & $      0.69$  & $       0.1$  & $        $  & $      -2.3$  & $        $  & $     0.574$  & $     0.880$  \\
R & $ 0.0000006$  & $    0.0007$  & $       2.0$  & $      0.06$  & $     0.009$  & $     0.004$  & $      0.03$  & $       0.1$  & $        $  & $       0.1$  & $        $  & $     0.043$  & $     0.084$  \\
\noalign{\smallskip}
1512bf & $ 0.6074343$  & $56100.9311$  & $      87.2$  & $      0.10$  & $     0.438$  & $     0.091$  & $      0.66$  & $       0.1$  & $       0.1$  & $      -2.1$  & $      -2.2$  & $     0.762$  & $     0.391$  \\
R+g & $ 0.0000002$  & $    0.0002$  & $       1.9$  & $      0.02$  & $     0.005$  & $     0.003$  & $      0.03$  & $       0.1$  & $       0.1$  & $       0.0$  & $       0.0$  & $     0.017$  & $     0.012$  \\
\noalign{\smallskip}
1613s & $ 1.1420695$  & $56511.1762$  & $      76.2$  & $      0.10$  & $     0.470$  & $     0.044$  & $      0.60$  & $       1.2$  & $       1.2$  & $      -2.2$  & $      -2.2$  & $     0.812$  & $     0.091$  \\
R+g & $ 0.0000024$  & $    0.0014$  & $       6.0$  & $      0.03$  & $     0.050$  & $     0.008$  & $      0.05$  & $       0.5$  & $       0.5$  & $       0.1$  & $       0.0$  & $     0.050$  & $     0.023$  \\
\noalign{\smallskip}
1613u & $ 0.5644902$  & $56787.5255$  & $      81.6$  & $      0.18$  & $     0.472$  & $     0.107$  & $      0.77$  & $       1.4$  & $       1.6$  & $      -2.4$  & $      -2.5$  & $     0.889$  & $     0.340$  \\
R+g & $ 0.0000003$  & $    0.0003$  & $       2.7$  & $      0.03$  & $     0.012$  & $     0.004$  & $      0.03$  & $       0.4$  & $       0.4$  & $       0.1$  & $       0.2$  & $     0.023$  & $     0.024$  \\
\noalign{\smallskip}
1615ag & $ 0.6806898$  & $55380.8645$  & $      85.8$  & $      0.17$  & $     0.410$  & $     0.079$  & $      0.76$  & $       2.2$  & $       1.9$  & $      -2.2$  & $      -2.4$  & $     0.772$  & $     0.358$  \\
R+g & $ 0.0000046$  & $    0.0010$  & $       3.3$  & $      0.04$  & $     0.013$  & $     0.005$  & $      0.05$  & $       1.0$  & $       0.8$  & $       0.3$  & $       0.1$  & $     0.026$  & $     0.029$  \\
\noalign{\smallskip}
1615v & $ 0.5594054$  & $54962.6621$  & $      73.7$  & $      0.10$  & $     0.457$  & $     0.099$  & $      0.71$  & $       0.2$  & $       0.1$  & $      -2.2$  & $      -2.9$  & $     0.784$  & $     0.410$  \\
R+g & $ 0.0000003$  & $    0.0007$  & $       1.4$  & $      0.02$  & $     0.012$  & $     0.004$  & $      0.03$  & $       0.1$  & $       0.1$  & $       0.1$  & $       0.3$  & $     0.029$  & $     0.031$  \\
\noalign{\smallskip}
1515ay & $ 0.4642873$  & $56138.1751$  & $      89.0$  & $      0.11$  & $     0.452$  & $     0.161$  & $      0.83$  & $       0.0$  & $       0.1$  & $      -2.2$  & $      -1.9$  & $     0.794$  & $     0.606$  \\
R+g & $ 0.0000001$  & $    0.0002$  & $       1.1$  & $      0.03$  & $     0.003$  & $     0.002$  & $      0.02$  & $       0.0$  & $       0.1$  & $       0.1$  & $       0.0$  & $     0.027$  & $     0.018$  \\
\noalign{\smallskip}
1615w & $ 1.4407151$  & $56530.5582$  & $      77.7$  & $      0.15$  & $     0.393$  & $     0.060$  & $      0.60$  & $       0.7$  & $       0.4$  & $      -2.9$  & $      -2.5$  & $     0.723$  & $     0.092$  \\
R+g & $ 0.0000024$  & $    0.0010$  & $       2.7$  & $      0.04$  & $     0.021$  & $     0.004$  & $      0.04$  & $       0.3$  & $       0.2$  & $       0.3$  & $       0.1$  & $     0.026$  & $     0.015$  \\
\noalign{\smallskip}
1615ao & $ 0.8954515$  & $56308.1400$  & $      77.6$  & $      0.24$  & $     0.368$  & $     0.129$  & $      0.94$  & $       2.1$  & $       1.9$  & $      -2.4$  & $      -2.5$  & $     0.729$  & $     0.273$  \\
R+g & $ 0.0000007$  & $    0.0005$  & $       0.8$  & $      0.06$  & $     0.007$  & $     0.005$  & $      0.02$  & $       1.1$  & $       1.0$  & $       0.3$  & $       0.2$  & $     0.038$  & $     0.020$  \\
\noalign{\smallskip}
1615u & $ 0.7777349$  & $56185.6219$  & $      82.4$  & $      0.16$  & $     0.297$  & $     0.038$  & $      0.59$  & $       1.4$  & $       1.8$  & $      -2.1$  & $      -2.2$  & $     0.554$  & $     0.726$  \\
R+g & $ 0.0000011$  & $    0.0011$  & $       4.2$  & $      0.07$  & $     0.034$  & $     0.006$  & $      0.05$  & $       0.7$  & $       0.8$  & $       0.1$  & $       0.0$  & $     0.041$  & $     0.191$  \\
\noalign{\smallskip}
1616cr & $ 0.5649690$  & $55972.1677$  & $      82.5$  & $      0.05$  & $     0.416$  & $     0.141$  & $      0.84$  & $       0.4$  & $       0.5$  & $      -2.3$  & $      -3.0$  & $     0.657$  & $     0.558$  \\
R+g & $ 0.0000002$  & $    0.0002$  & $       0.9$  & $      0.01$  & $     0.005$  & $     0.002$  & $      0.02$  & $       0.2$  & $       0.1$  & $       0.2$  & $       0.2$  & $     0.021$  & $     0.025$  \\
\noalign{\smallskip}

\hline
\end{tabular}
\end{table*}
\begin{table*}
\contcaption{}
\begin{tabular}{@{}l *{13}{r} }
ID & $P\,\mathrm{(d)}$ & $t_0\,\mathrm{(BMJD_{tdb})}$ & $i\,(\degree)$ & $q$ & $\langle r_1 \rangle$ & $\langle r_2 \rangle$ & $T_2/T_1$ & $\mathrm{absorb}_R$ & $\mathrm{absorb}_{g^\prime}$ & $\log(f_{R})$ & $\log(f_{g^\prime})$ & fill & $\rho\,(\rho_\mathrm{\sun})$ \\
Band & & & & & & & & & & & & & \\
\hline
\noalign{\smallskip}
1617n & $ 2.3367776$  & $55591.4345$  & $      87.3$  & $      0.21$  & $     0.258$  & $     0.040$  & $      0.63$  & $       2.6$  & $       1.5$  & $      -2.1$  & $      -2.4$  & $     0.501$  & $     0.117$  \\
R+g & $ 0.0000052$  & $    0.0027$  & $       2.1$  & $      0.07$  & $     0.012$  & $     0.003$  & $      0.04$  & $       0.8$  & $       0.5$  & $       0.1$  & $       0.1$  & $     0.025$  & $     0.013$  \\
\noalign{\smallskip}
1617m & $ 3.7728999$  & $56584.2146$  & $      87.8$  & $      0.24$  & $     0.206$  & $     0.056$  & $      0.68$  & $       0.6$  & $       0.7$  & $      -2.1$  & $      -2.2$  & $     0.410$  & $     0.085$  \\
R+g & $ 0.0000083$  & $    0.0010$  & $       1.5$  & $      0.15$  & $     0.005$  & $     0.002$  & $      0.03$  & $       0.5$  & $       0.4$  & $       0.1$  & $       0.1$  & $     0.046$  & $     0.011$  \\
\noalign{\smallskip}
1619l & $ 1.1599993$  & $56560.2644$  & $      82.7$  & $      0.14$  & $     0.385$  & $     0.061$  & $      0.67$  & $       1.2$  & $       3.6$  & $      -2.2$  & $      -2.3$  & $     0.704$  & $     0.153$  \\
R+g & $ 0.0000017$  & $    0.0016$  & $       4.2$  & $      0.04$  & $     0.022$  & $     0.006$  & $      0.04$  & $       0.4$  & $       0.8$  & $       0.1$  & $       0.3$  & $     0.029$  & $     0.021$  \\
\noalign{\smallskip}
1521ct & $ 1.1724964$  & $56907.7345$  & $      83.2$  & $      0.17$  & $     0.282$  & $     0.093$  & $      0.92$  & $       1.2$  & $       2.2$  & $      -2.3$  & $      -2.9$  & $     0.528$  & $     0.373$  \\
R+g & $ 0.0000013$  & $    0.0005$  & $       1.4$  & $      0.09$  & $     0.009$  & $     0.004$  & $      0.02$  & $       0.8$  & $       1.1$  & $       0.1$  & $       0.3$  & $     0.046$  & $     0.044$  \\
\noalign{\smallskip}
1621ax & $ 1.0181525$  & $56741.7492$  & $      83.1$  & $      0.17$  & $     0.401$  & $     0.029$  & $      0.57$  & $       2.9$  & $       2.2$  & $      -2.2$  & $      -2.2$  & $     0.758$  & $     0.171$  \\
R+g & $ 0.0000044$  & $    0.0017$  & $       4.8$  & $      0.04$  & $     0.028$  & $     0.006$  & $      0.07$  & $       0.9$  & $       1.1$  & $       0.0$  & $       0.1$  & $     0.031$  & $     0.027$  \\
\noalign{\smallskip}
1521cm & $ 0.6854774$  & $56068.9363$  & $      80.0$  & $      0.22$  & $     0.381$  & $     0.110$  & $      0.78$  & $       0.7$  & $       0.8$  & $      -2.9$  & $      -2.6$  & $     0.740$  & $     0.428$  \\
R+g & $ 0.0000002$  & $    0.0002$  & $       1.0$  & $      0.04$  & $     0.008$  & $     0.003$  & $      0.02$  & $       0.3$  & $       0.2$  & $       0.3$  & $       0.2$  & $     0.025$  & $     0.027$  \\
\noalign{\smallskip}
1622by & $ 0.7486683$  & $55718.0157$  & $      85.6$  & $      0.16$  & $     0.424$  & $     0.076$  & $      0.81$  & $       3.5$  & $       3.1$  & $      -2.4$  & $      -2.9$  & $     0.791$  & $     0.270$  \\
R+g & $ 0.0000015$  & $    0.0010$  & $       3.6$  & $      0.03$  & $     0.017$  & $     0.005$  & $      0.04$  & $       0.9$  & $       0.8$  & $       0.3$  & $       0.3$  & $     0.024$  & $     0.025$  \\
\noalign{\smallskip}
1522cc & $ 0.5717853$  & $56641.9749$  & $      81.2$  & $      0.18$  & $     0.472$  & $     0.078$  & $      0.69$  & $       1.4$  & $       1.2$  & $      -2.4$  & $      -2.6$  & $     0.894$  & $     0.331$  \\
R+g & $ 0.0000003$  & $    0.0003$  & $       3.1$  & $      0.03$  & $     0.013$  & $     0.004$  & $      0.04$  & $       0.3$  & $       0.3$  & $       0.1$  & $       0.2$  & $     0.018$  & $     0.023$  \\
\noalign{\smallskip}
1622aa & $ 0.7661291$  & $57136.1682$  & $      84.7$  & $      0.10$  & $     0.410$  & $     0.060$  & $      0.75$  & $       0.6$  & $       1.4$  & $      -2.4$  & $      -2.9$  & $     0.716$  & $     0.302$  \\
R+g & $ 0.0000038$  & $    0.0014$  & $       4.0$  & $      0.03$  & $     0.022$  & $     0.006$  & $      0.05$  & $       0.5$  & $       0.7$  & $       0.2$  & $       0.3$  & $     0.026$  & $     0.038$  \\
\noalign{\smallskip}
1622bt & $ 0.6884160$  & $56746.2432$  & $      79.2$  & $      0.17$  & $     0.425$  & $     0.071$  & $      0.70$  & $       1.7$  & $       1.4$  & $      -2.3$  & $      -2.5$  & $     0.795$  & $     0.314$  \\
R+g & $ 0.0000004$  & $    0.0003$  & $       2.0$  & $      0.02$  & $     0.012$  & $     0.003$  & $      0.03$  & $       0.4$  & $       0.3$  & $       0.0$  & $       0.2$  & $     0.017$  & $     0.023$  \\
\noalign{\smallskip}
1723aj & $ 1.1088064$  & $56733.1351$  & $      85.6$  & $      0.11$  & $     0.460$  & $     0.042$  & $      0.52$  & $       0.5$  & $       0.5$  & $      -2.4$  & $      -3.1$  & $     0.818$  & $     0.100$  \\
R+g & $ 0.0000009$  & $    0.0004$  & $       3.3$  & $      0.02$  & $     0.011$  & $     0.003$  & $      0.05$  & $       0.2$  & $       0.2$  & $       0.0$  & $       0.2$  & $     0.022$  & $     0.006$  \\
\noalign{\smallskip}
\hline
\end{tabular}
\end{table*}

\clearpage

\begin{figure*}
\includegraphics{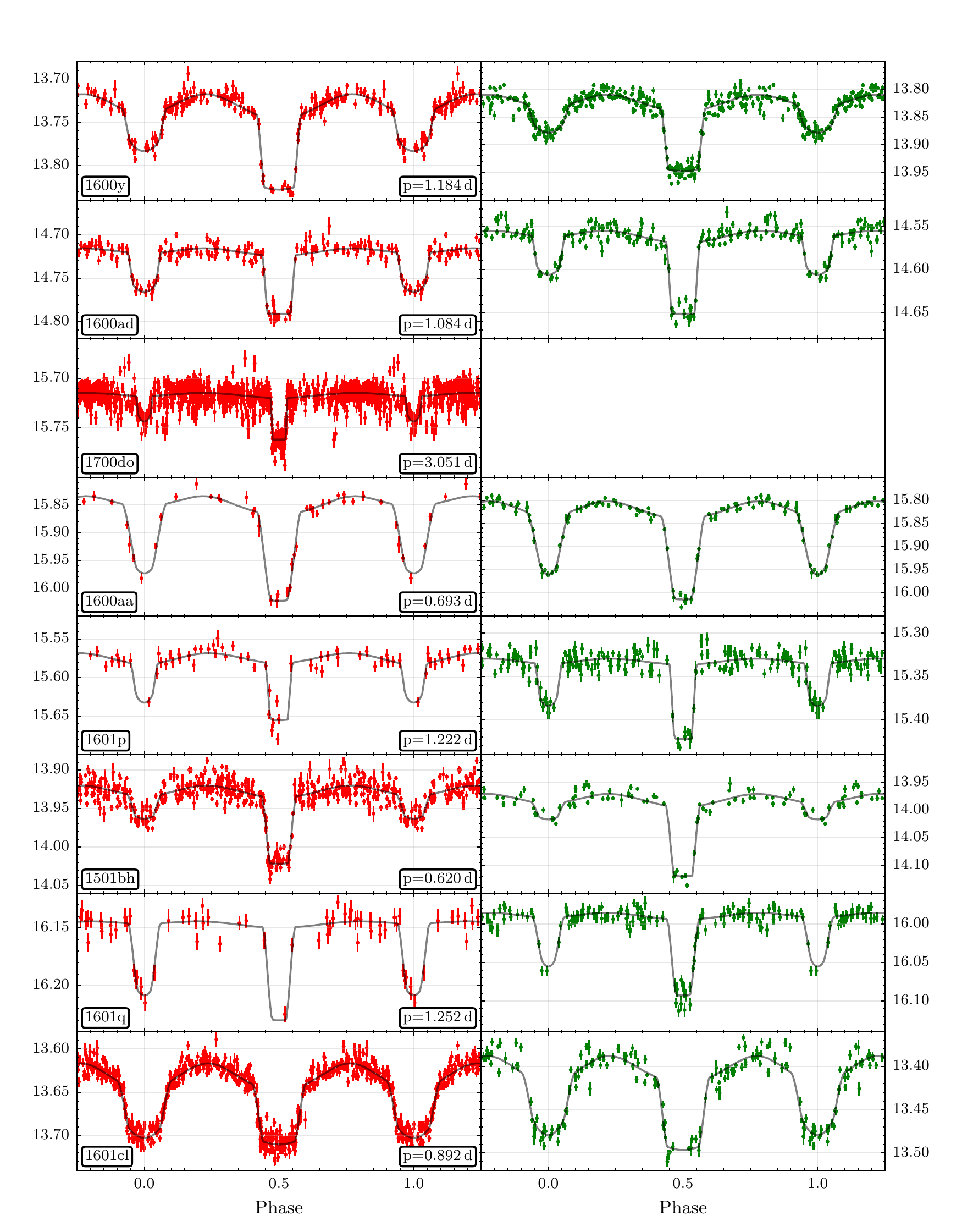}
\caption{The PTF lightcurves in $R$ (left) and $g^\prime$ (right) with the best model over-plotted (see Table~\ref{tab:lcstats}).}
\label{fig:LCs}
\end{figure*}

\begin{figure*}
\includegraphics{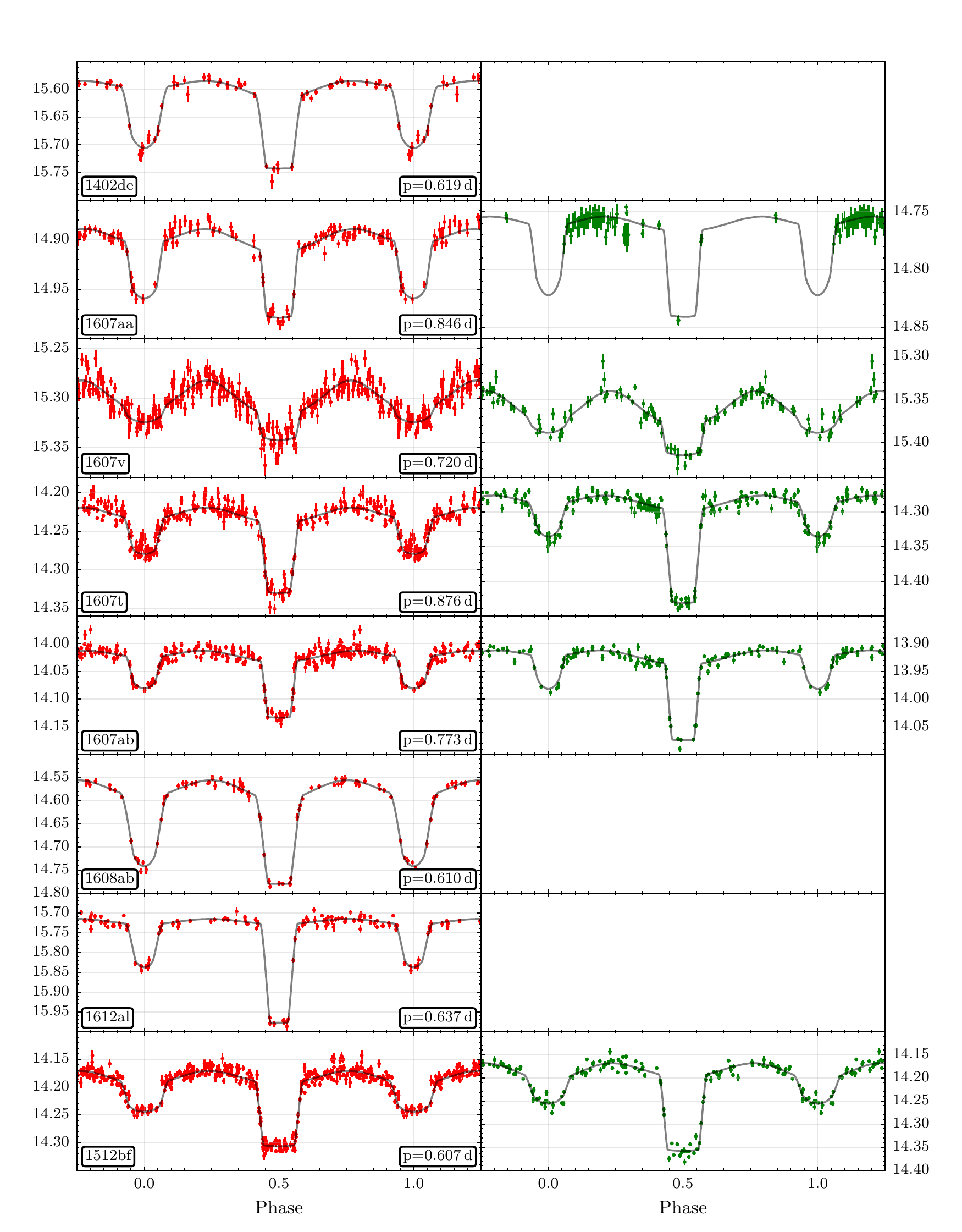}
\contcaption{}
\end{figure*}

\begin{figure*}
\includegraphics{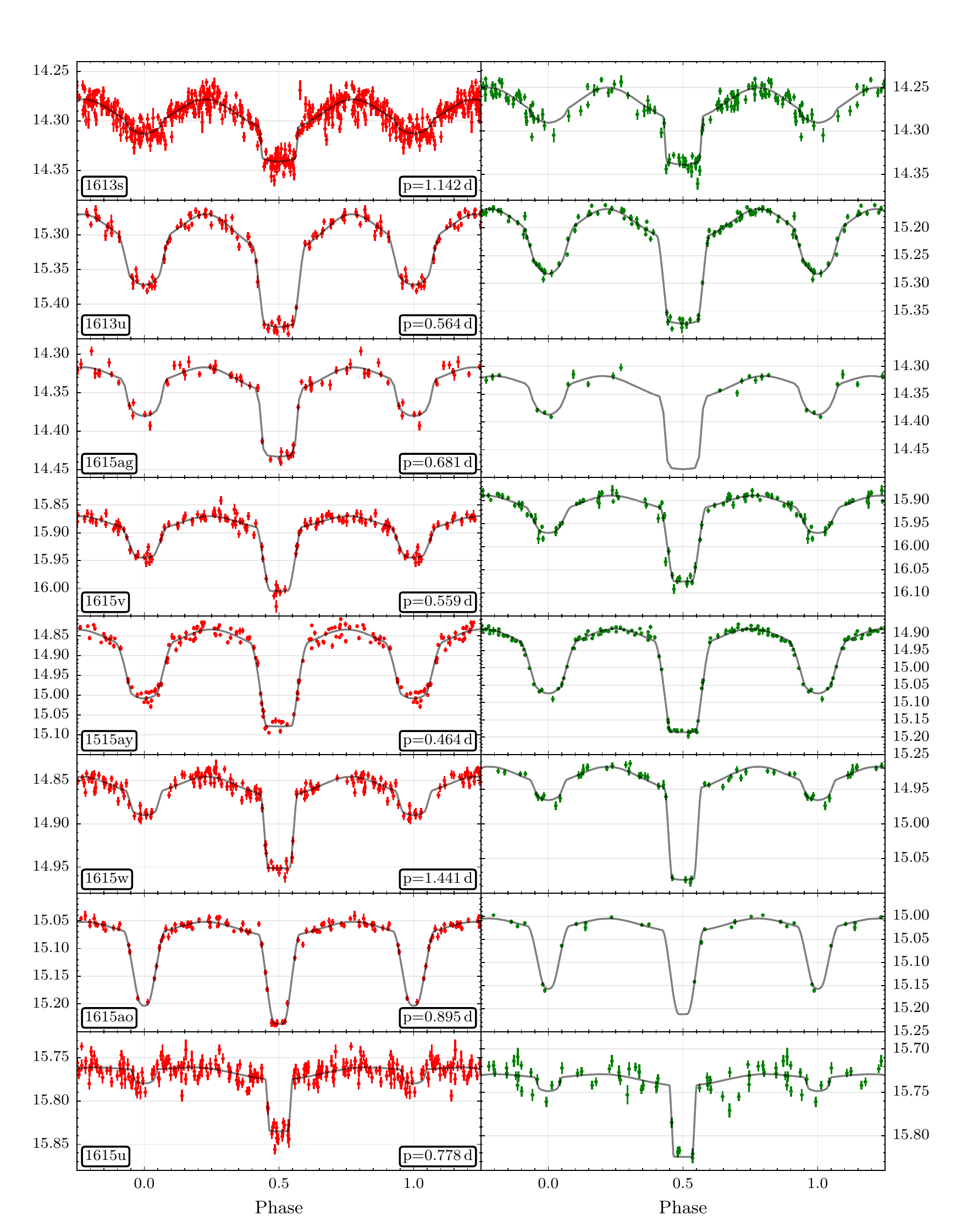}
\contcaption{}
\end{figure*}

\begin{figure*}
\includegraphics{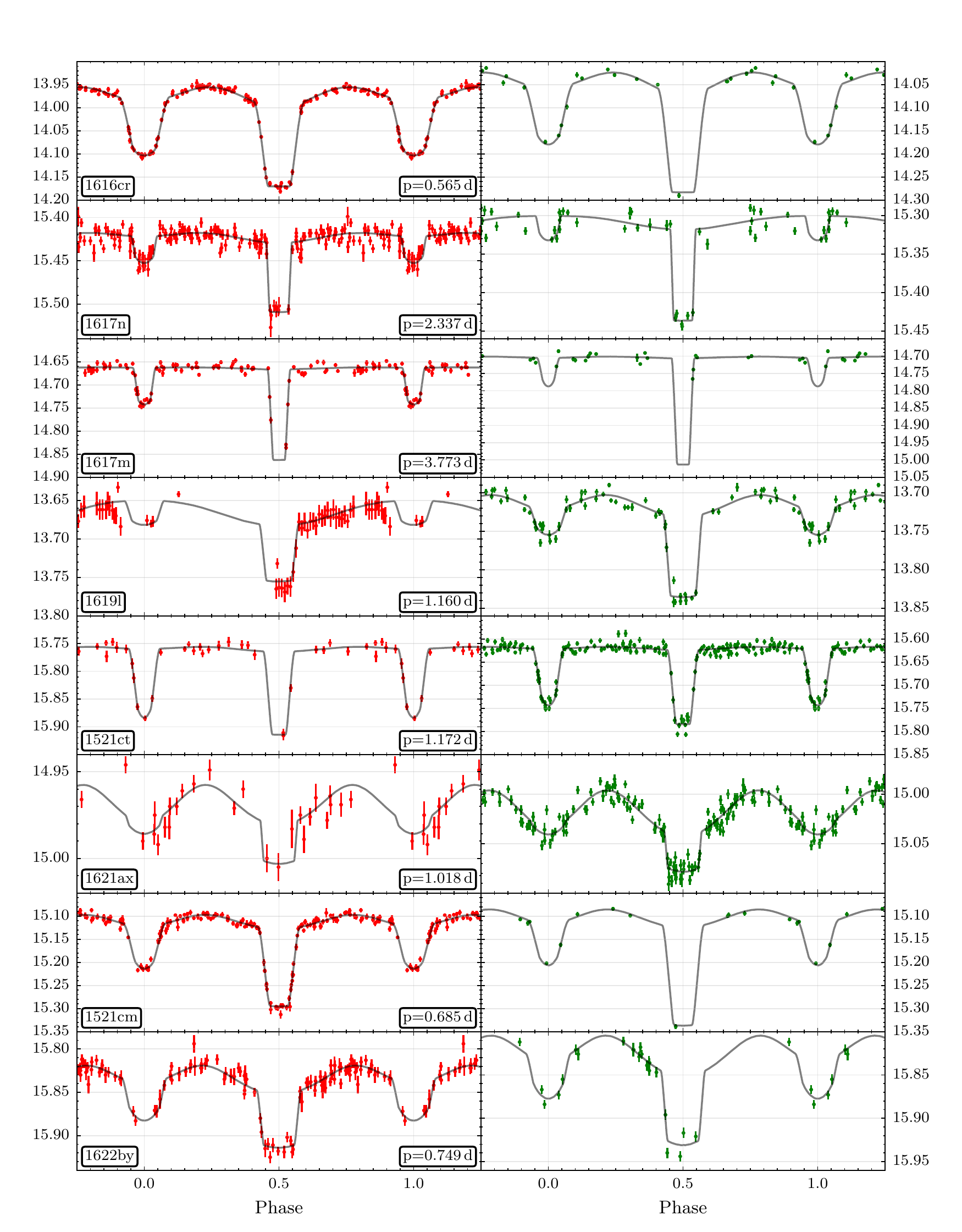}
\contcaption{}
\end{figure*}

\begin{figure*}
\includegraphics{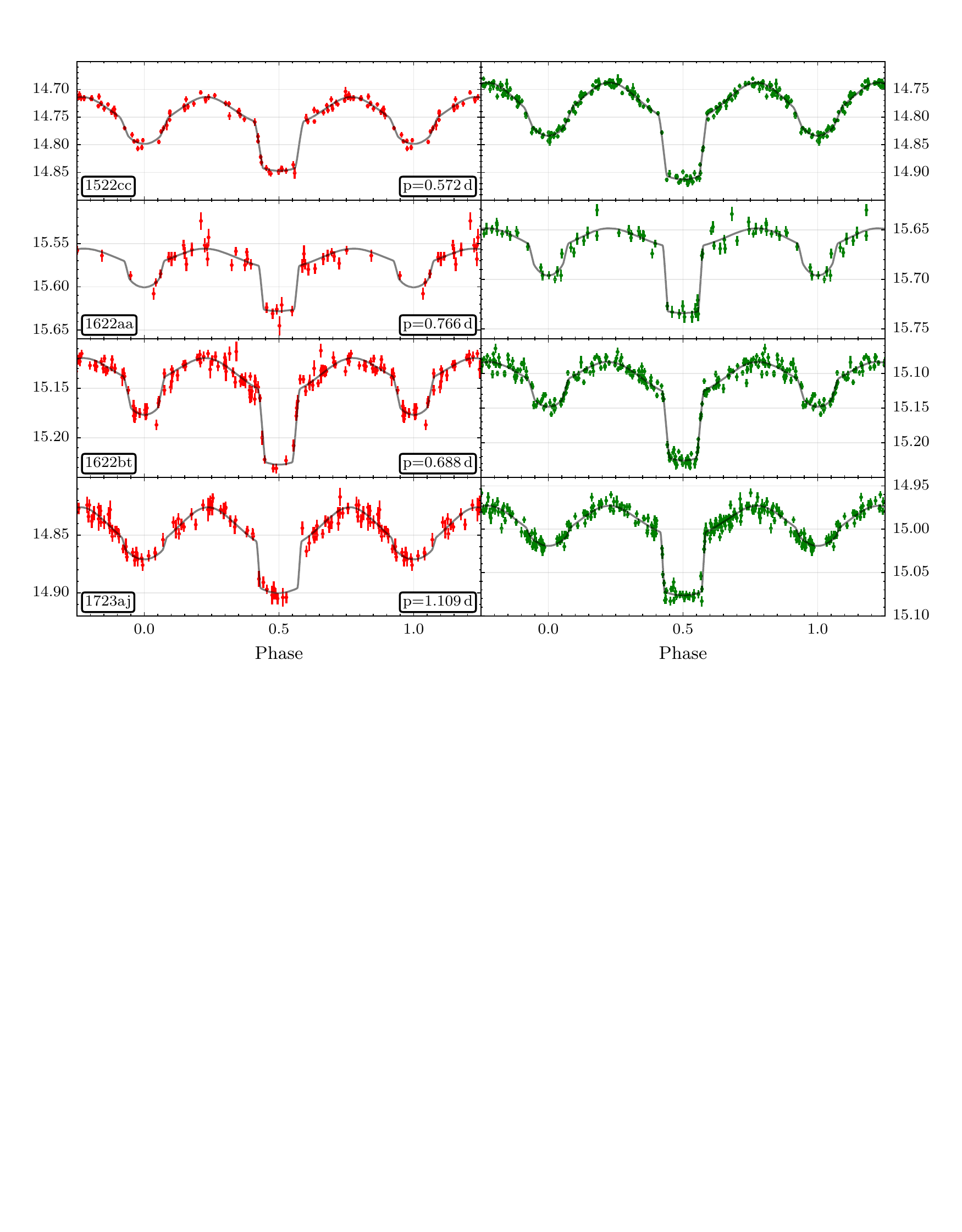}
\contcaption{}
\end{figure*}

\begin{figure*}
\includegraphics{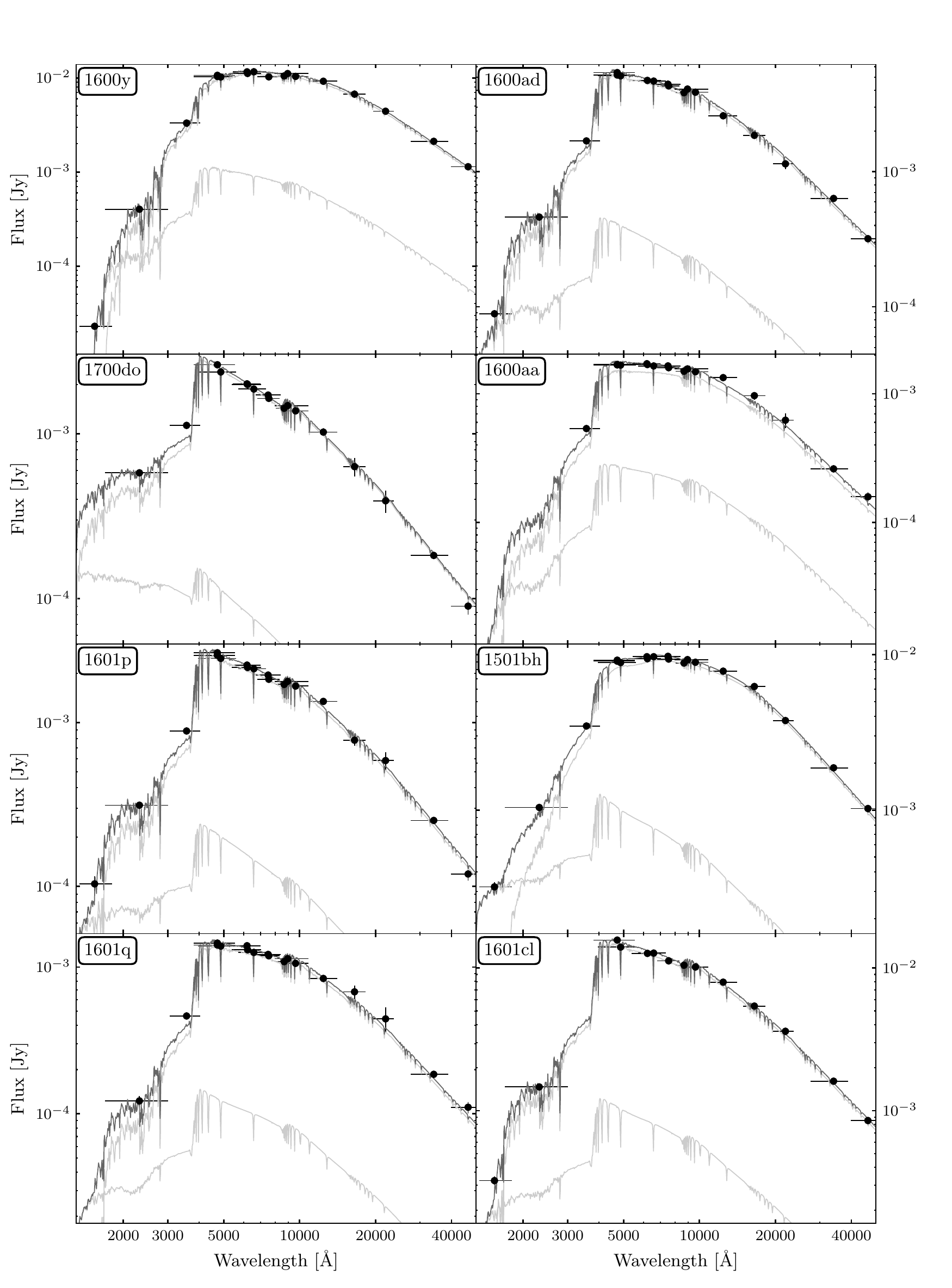}
\caption{The spectral energy distribution and the best-fitting model spectra (see Table~\ref{tab:SEDfit}). The grey lines show the SED of the A/F-star and pre-He-WD. The black line shows the sum of both components. The A/F-star dominates the SED over the whole wavelength range, except in the far-UV in some of the cases.}
\label{fig:SEDcurves}
\end{figure*}

\begin{figure*}
\includegraphics{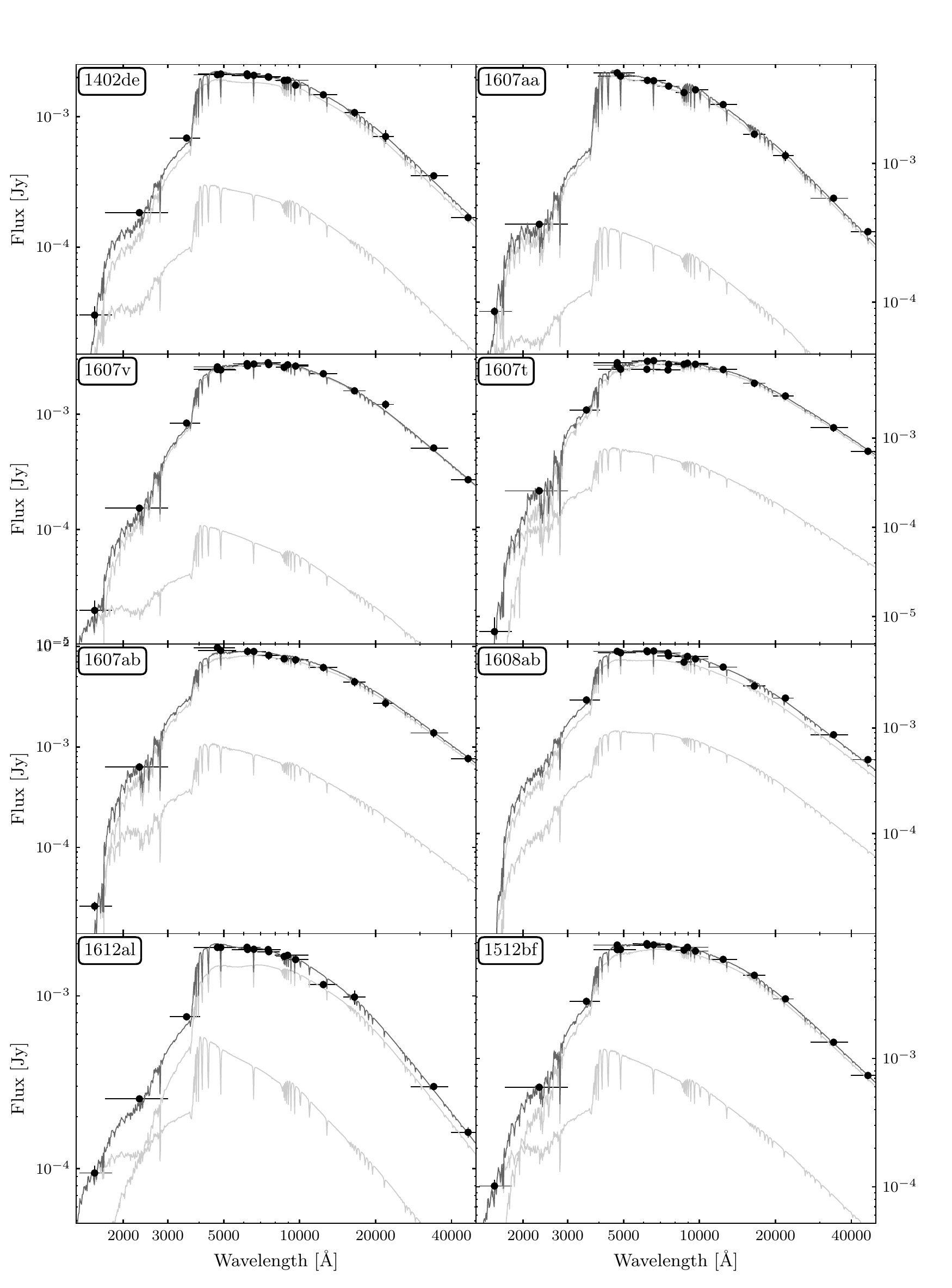}
\contcaption{}
\end{figure*}

\begin{figure*}
\includegraphics{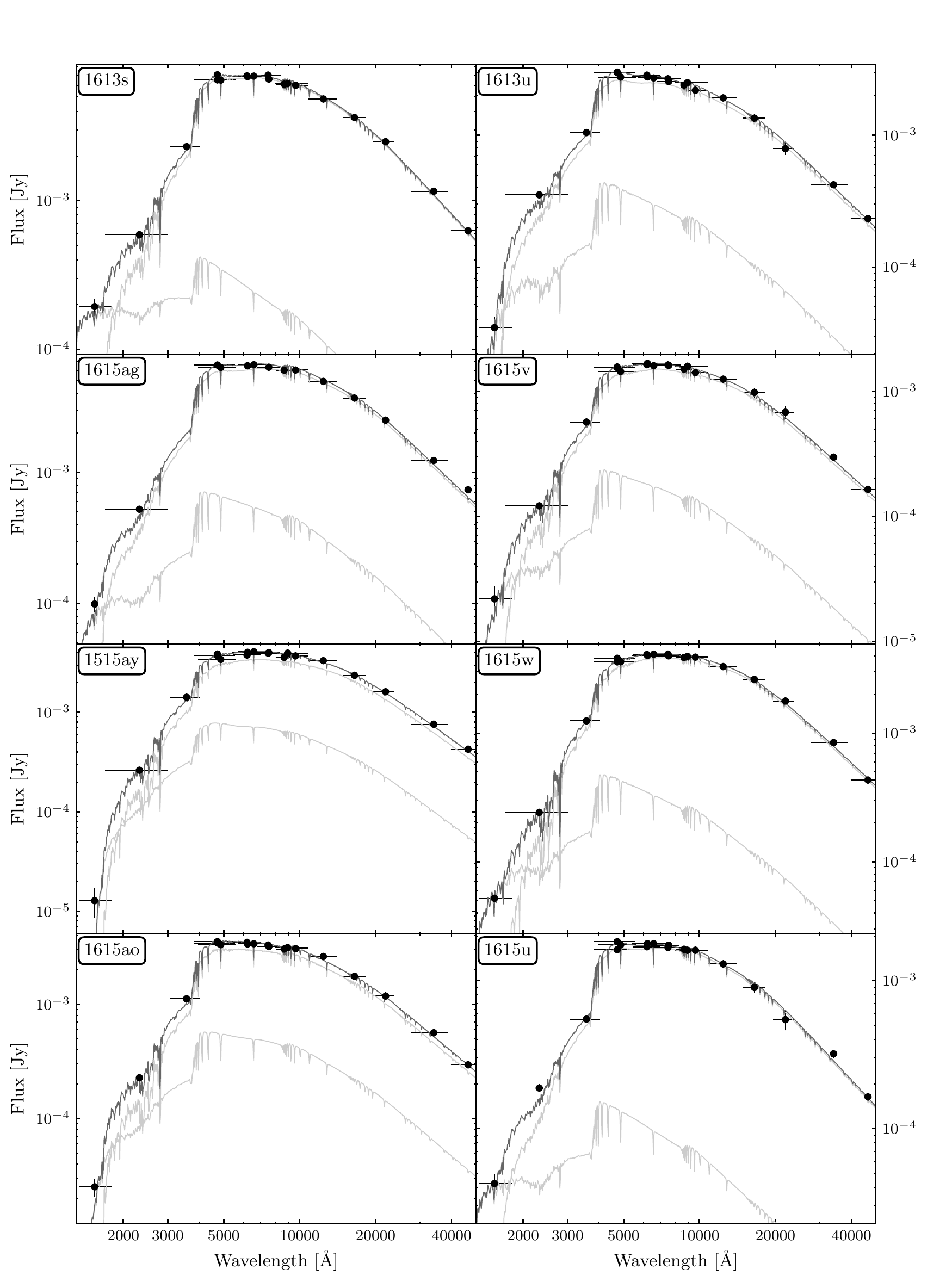}
\contcaption{}
\end{figure*}

\begin{figure*}
\includegraphics{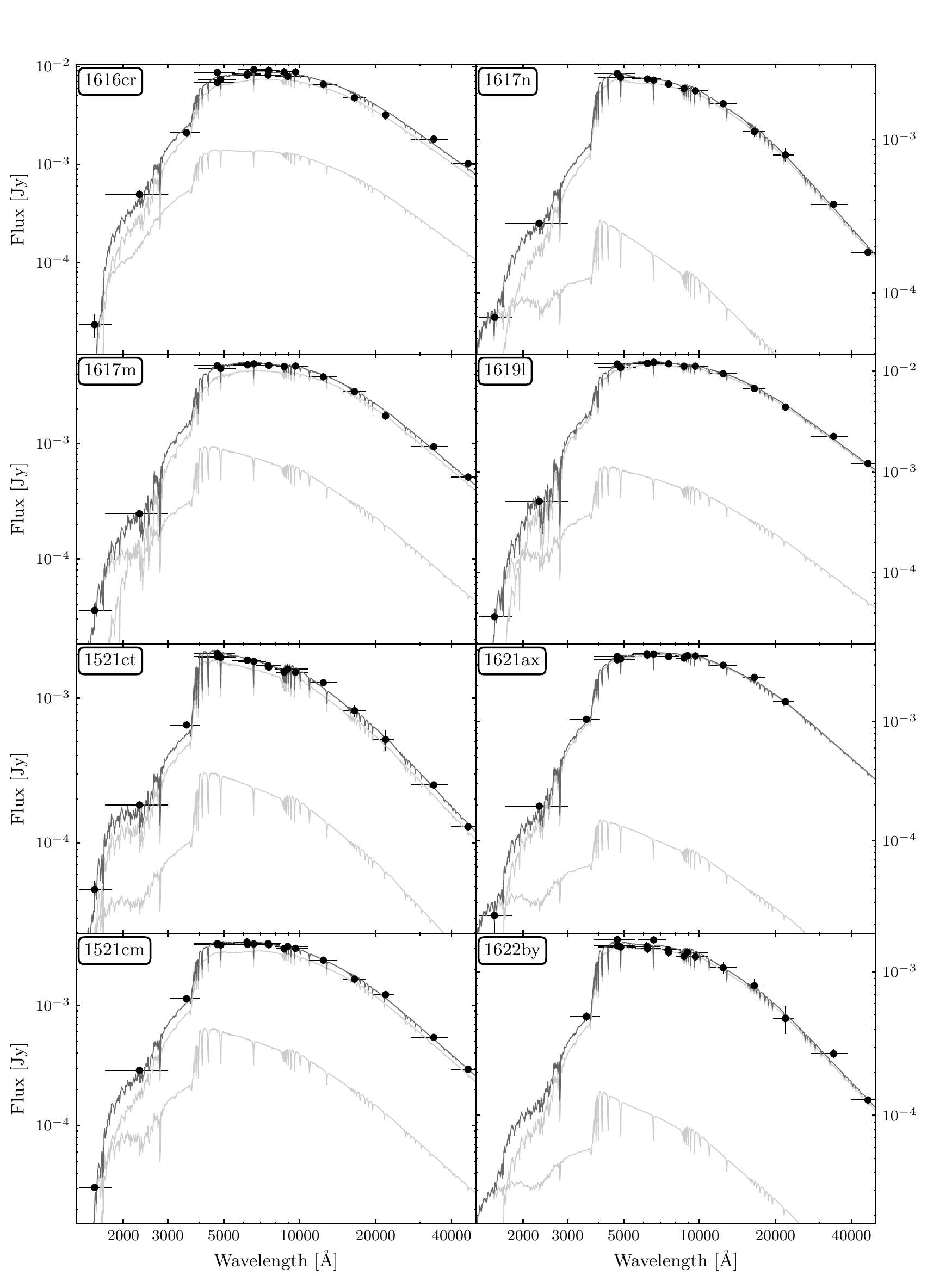}
\contcaption{}
\end{figure*}

\begin{figure*}
\includegraphics{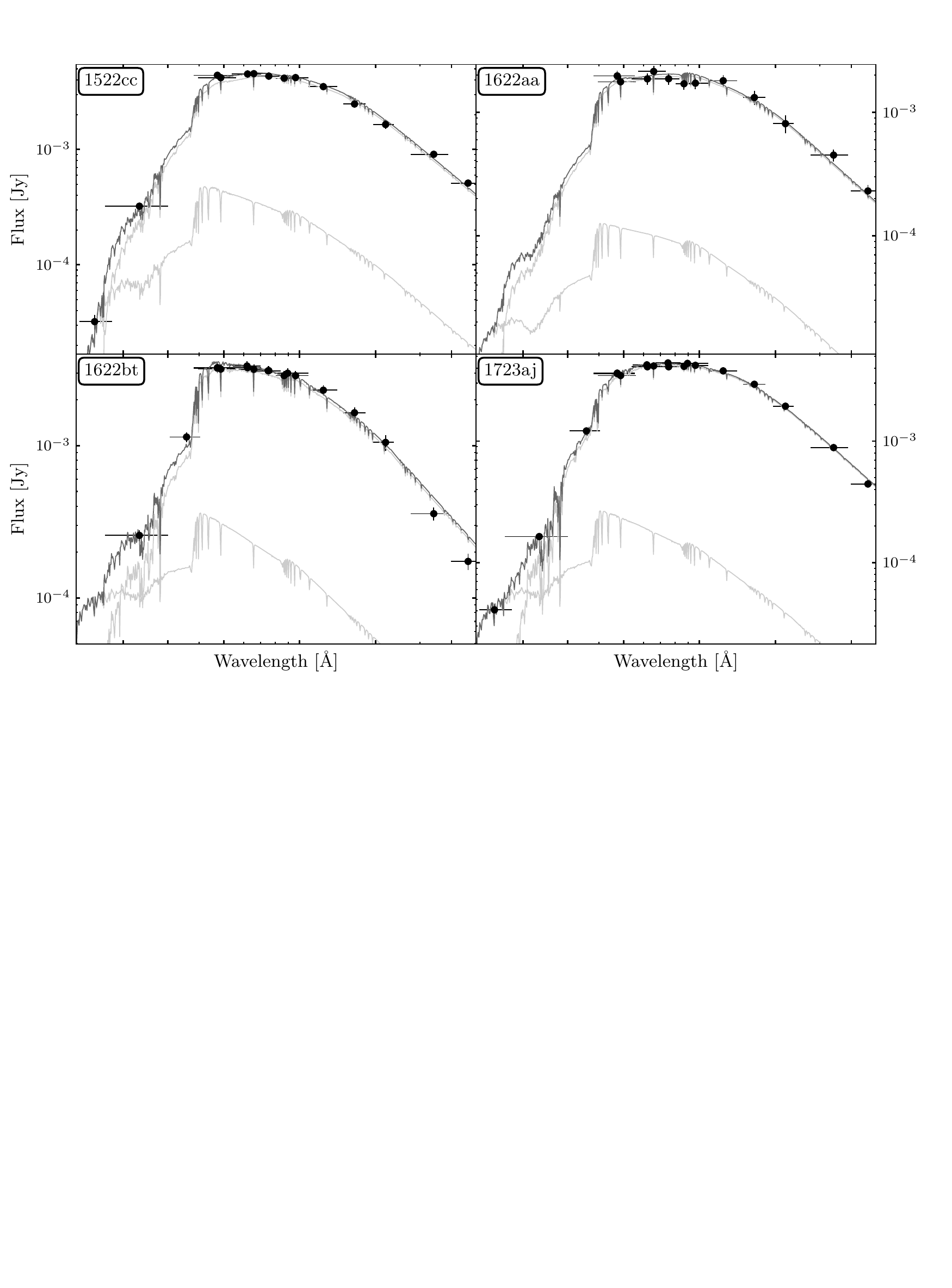}
\contcaption{}
\end{figure*}

\begin{figure*}
\mbox{\includegraphics[width=1.0\textwidth,height=0.9\textheight]{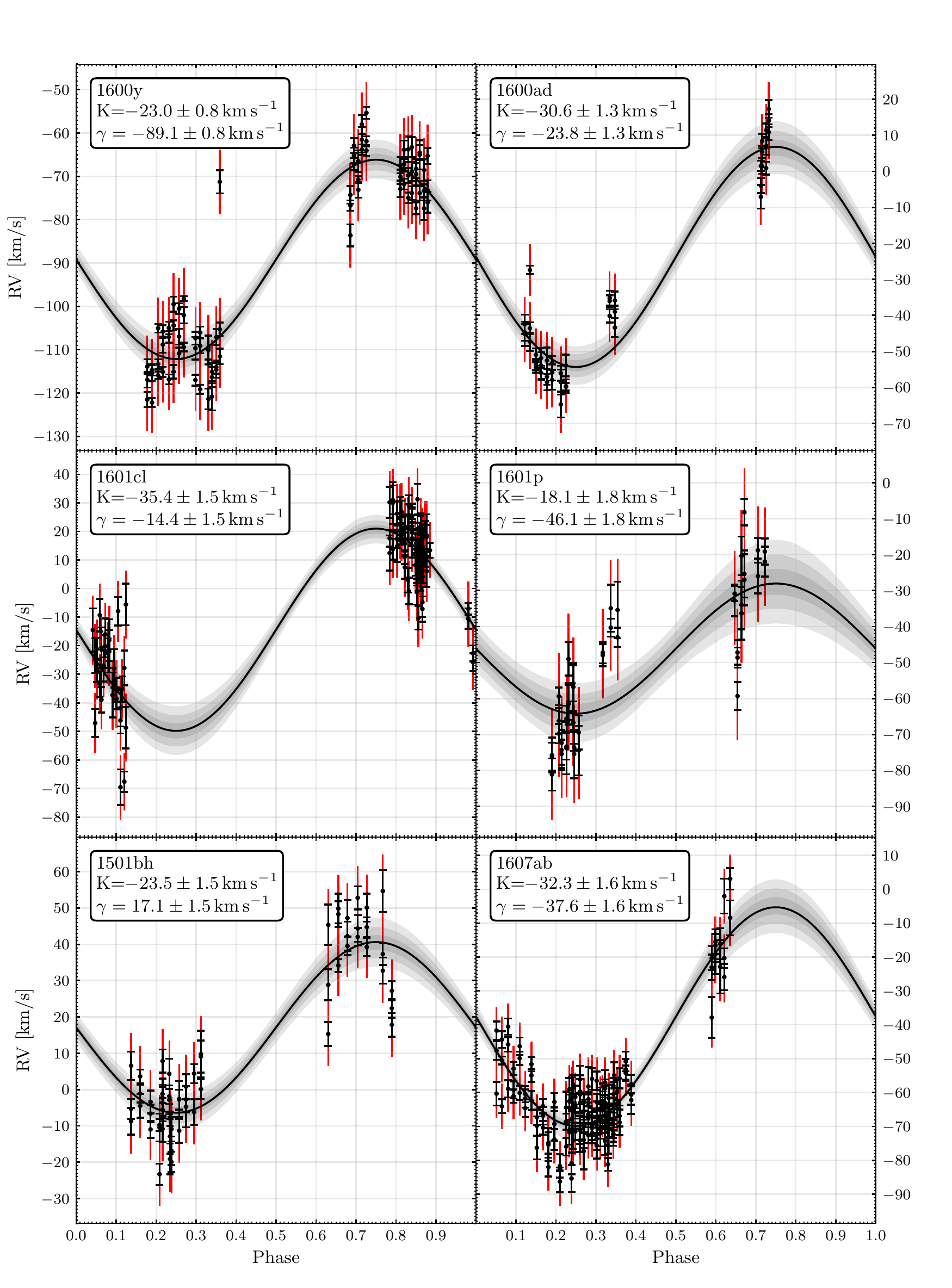}}
\caption{The radial velocity measurements with the INT and the best fitting model. Black errorbars show the estimated uncertainty from the cross-correlation procedure, while the red errorbars show the uncertainties required to account for all residual variance. The shaded grey contours show the 1, 2 and 3 standard deviation intervals of model, obtained using the larger uncertainties.}
\label{fig:RVcurves}
\end{figure*}

\begin{figure*}
\mbox{\includegraphics[width=1.0\textwidth,height=0.9\textheight]{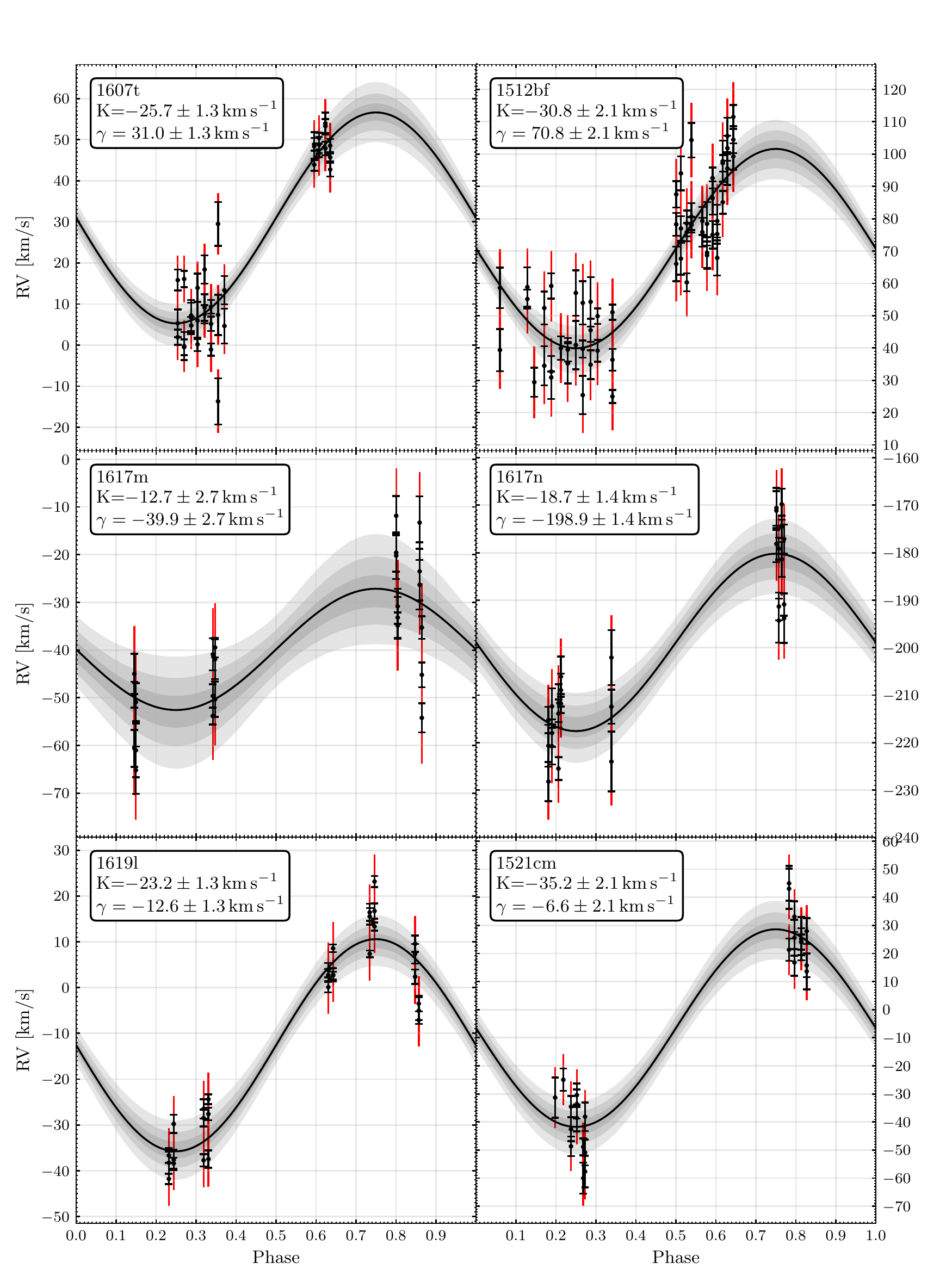}}
\contcaption{}
\end{figure*}


\label{lastpage}
\end{document}